June 26, 2011 (with Summary)

# MIXED LAYER MESOSCALES FOR OGCMS:

## Model development and assessment with T/P, WOCE and Drifter data

## Tracer fluxes, kinetic energy, diffusivity, dynamic tapering, new bolus velocity

V.M. Canuto[1,2], M.S. Dubovikov[1,3] and A. Leboissetier[1,4]

[1]NASA, Goddard Institute for Space Studies, 2880 Broadway, New York, NY, 10025
[2]Dept. Applied of Phys. and Math., Columbia University, New York, NY, 10027
[3]Center for Climate System research, Columbia Univ., New York, NY, 10025
[4]Dept. of Earth, Atmospheric and Planetary Sciences, MIT, Cambridge, Mass., 02138

# Abstract

In this paper we present a model for mixed layer (ML) mesoscale fluxes of an arbitrary tracer in terms of the resolved fields (mean tracer and mean velocity). The treatment of an arbitrary tracer, rather than only buoyancy, is necessary since OGCMs (ocean global circulation models) time step T, S, CO2, etc and not buoyancy. The particular case of buoyancy is used to assess the model results.

The paper contains three parts: derivation of the results, discussion of the results and assessment of the latter using, among others, WOCE, T/P and Drifter data.

**Derivation.** To construct the mesoscale fluxes, we first solve the ML mesoscale dynamic equations for the velocity and tracer mesoscale fields. The goal of the derivation is to emphasize the different treatments of the non-linear terms in the adiabatic vs. diabatic ocean (deep ocean vs. mixed layer).

**Results.** We derive analytic expressions for the following variables: **a)** vertical and horizontal mesoscale fluxes of an arbitrary tracer, **b)** mesoscale diffusivity in terms of the eddy kinetic energy, **c)** surface value of the latter in terms of the vertical mesoscale buoyancy flux together with a model for the z-profile of the mesoscale kinetic energy, **d)** tapering function T(z) in terms of the large scale variables; T(z) vanishes at the surface and tends to unity below the ML where the stream function smoothly connects with the deep ocean GM form, **e)** new eddy induced velocity. Thus far, tracers were described by only one eddy induced velocity (curl of the stream function); for tracers other than buoyancy, we show there is an additional bolus velocity originating not from the stream function but from the residual flux which exhibits a skew component.

**Assessment.** The model assessment includes the following items: **a)** the vertical flux naturally vanishes at the ocean surface, as physically required**, b)** the second z-derivative of the buoyancy flux is negative, implying re-stratification, in agreement with eddy resolving simulations, **c)** the predicted surface eddy kinetic energy compares well with the T/P-Jason-1 altimetry data in both intensity and geographical distribution, **d)** the predicted z-profile of the eddy kinetic energy compares well with WOCE data, **e)** the model predicts both the z-profile and the surface values of the mesoscale diffusivity, **f)** the latter is in accord with the Global Drifter and T/P data.

# 1. Introduction

The relevance of ocean mesoscales, M, and the inability of coarse resolution OGCMs to resolve them, prompted the search for a parameterization of M in terms of the resolved fields. For completeness, we recall some of the key features of oceanic M[1]. They are ubiquitous, large structures of the order of the Rossby deformation radius, lifetimes of the order of several months and a kinetic energy that on average exceeds that of the mean flow (Sharffenberg and Stammer, 2010, Fig.7; Munk, 2011). Observations (Richardson, 1993) show that "M are water-mass anomalies with nearly circular flow around their centers which move through the background water at speeds and directions inconsistent with background flow", that is, M move in a mean flow with their own velocity, often referred to as drift velocity (Henson and Thomas, 2008). Mesoscales form coherent, almost axi-symmetric structures in which all the features at different depths are highly correlated (Provenzale, 1999). Finally, they re-stratify the mixed layer (Nurser et al., 2000; Oschlies, 2002; Hosegood et al., 2008), thus opposing the de-stratification by vertical turbulent mixing. Destratification affects the ML depth, a feature closely related to the ocean's ability to absorb heat and CO2 which are of primary relevance to climate studies (e.g., Sarmiento et al., 1998; Hansen et al., 2011).

The task of constructing M parameterizations is different in the *deep ocean* and *mixed layer* (ML), the former being largely adiabatic, while the latter is largely diabatic. In the adiabatic regime, mesoscales are generally parameterized with the GM model (Gent and McWilliams, 1990) which Gille and Davis (1999) showed to possess a skill of 40% compared to the much lower skill of previous models that parameterized M with a horizontal diffusivity. The mesoscale parameterization in the deep ocean has been traditionally described using the RMT formalism (residual mean theory, Andrew and McIntyre, 1976; Treguier et al., 1997; Ferreira et al., 2006; Ferrari et al., 2008), in which one separates the buoyancy fluxes in *along* and *across* isopycnal surfaces, the first being described by a stream function $\Psi$ and the latter by a residual flux $\mathbf{F}_r$. The measured smallness of the deep ocean diapycnal fluxes (Ledwell et al., 1998; 2011) reduces the problem to the parameterization of essentially one variable, the stream function $\Psi$ which contributes an advective velocity (referred to as eddy induced velocity or bolus velocity) which GM (Gent, 2011) parameterized in terms of the down-gradient of the isopycnal slopes. Physically, this corresponds to a flattening of the isopycnal slopes so as to release mean potential energy via baroclinic instabilities.

Mixed layer mesoscales are difficult to parameterize because some of the advantages of the deep ocean regime are no longer present. As pointed out by Killworth (2005), the large turbulent vertical diffusivities make the ML strongly diabatic which means that the residual flux is no longer negligible as it was in the deep ocean and the M parameterization is consequently more complex. As a consequence, ML water parcels no longer move along isopycnals but along horizontal surfaces. Consequently, the RMT separation of fluxes is no longer as advantageous as it was in the deep ocean; instead, one should work in terms of vertical-horizontal fluxes. Second, in the ML, it is not possible to use the GM model as suggested for the deep ocean ($\kappa_M$ is the mesoscale diffusivity, $\mathbf{s}$ is the slope of the isopycnals and $\mathbf{e}_z$ is the unit vertical vector):

---

[1] W. Munk (2011) has recently called the discovery of mesoscales *The mesoscale revolution*.

$$\mathbf{\Psi}_{GM} = -\kappa_M \mathbf{s} \times \mathbf{e}_z \qquad (1a)$$

since (1a) does not satisfy the condition $\mathbf{\Psi}(0) = 0$ that is needed to ensure that $w^+ = (\nabla_H \times \mathbf{\Psi}_H) \cdot \mathbf{e}_z$ vanishes at z=0. A first suggested amendment was to introduce in (1a) an empirical function smoothly tapered to vanish at z=0 (Griffies, 2004). However, the arbitrariness in the choice of the *tapering functions* resulted in widely varying characteristics of the simulated flows (Gnanadesikan et al., 2007; Ferrari et al., 2008). This, in turned, elicited less ad hoc suggestions such as those proposed by Aiki et al. (2004, A4) and Ferrari et al. (2008, 2010, F8,10). Though superior to the tapering procedures, these models are still not complete, as discussed elsewhere (Canuto and Dubovikov, 2011, CD11)[2]. On the other hand, some of the features of the present model are as follows:

*a)* *dynamically based tapering*. A physically viable ML mesoscale parameterization must satisfy not only the surface boundary condition $\mathbf{\Psi}(0) = 0$ but it must also merge into the $\mathbf{\Psi}$ of the deep ocean which we take to be of the GM type (1a). It would be a desirable feature for a parameterization to satisfy these two conditions without the need of ad-hoc tapering. The model we present below yields the following new ML stream function:

$$\mathbf{\Psi} = -\kappa_M T(z) \mathbf{s} \times \mathbf{e}_z \qquad (1b)$$

which depends on the mesoscale diffusivity-isopycnal slope as the GM model (1a). However, it now contains a *dynamical tapering function* T(z) that will be shown to vanish at the surface (making the stream function vanish) and to become unity at the bottom of the ML, where (1b) smoothly merges into (1a). The function T(z) is given by the model in terms of the large scale variables, Eq.(39); it is not a universal function, but location dependent (Fig.5).

*b)* *residual flux $\mathbf{F}_r$*. The parameterization of the non-zero ML residual flux $\mathbf{F}_r$ poses new problems. Since in the deep ocean it is assumed to be zero, no parameterization exists to rely upon; $\mathbf{F}_r$ was only briefly discussed in F10 where it was considered to be diffusive. In this work, we show that $\mathbf{F}_r$ is diffusive but it has also a new feature, a skew component that gives rise to an additional bolus velocity $\mathbf{u}_{**}$ to be added to $\mathbf{u}^+ = \nabla \times \mathbf{\Psi}$. This means that both the stream function and the residual flux contribute an eddy induced velocity and thus, the traditional assumption of a single eddy induced velocity is no longer valid (see Figs.6).

---

[2] As summarized in CD11: a) the F8 eddy induced velocity does not entail ML mesoscale re-stratification which is known to exist, b) the A4,F10 eddy induced velocities yield re-stratification but, at the lowest order in the smallness parameter h/H, neither model exhibits a dependence of the ML re-stratification on $N^2(z)$ and, in that respect, A4, F10 are not different, c) A4,F10 do not provide the residual flux $\mathbf{F}_r$. CD11 showed that in A4-F10, $\mathbf{F}_r$ is diffusive with a mesoscale diffusivity that vanishes at the bottom of the ML, as expected, however, d) the surface diffusivity in such models is roughly h/H times smaller than the standard value of about $10^3 m^2 s^{-1}$ (h and H are the depths of the ML and the ocean, respectively).

**c)** *mesoscale diffusivity* $\kappa_M$. A ML mesoscale parameterization must also provide the mesoscale diffusivity $\kappa_M$ that appears in relations (1a,b). At present, there are two approaches toward this variable. One employs heuristic expressions such as[3]:

$$\kappa_M: \quad \text{const}, \quad \sim N^2, \quad \sim r_d^2 f Ri^{-1/2}, \quad (1-3)10^3 m^2 s^{-1}, \quad \ell^2 N |s| \tag{1c}$$

which are usually assessed using OGCMs. A second, data based approach employs results from T/P and Global Drifters and on this basis several authors (Stammer, 1998; Zhurbas and Oh, 2003, 2004; Rupolo, 2007; Sallee' et al., 2008a,b; 2010) proposed the following relation:

$$\kappa_M(0) = C\ell K(0)^{1/2} \quad , \quad C=O(1) \tag{1d}$$

where K(z) is the mesoscale eddy kinetic energy and K(0) its surface value. Being based on measured data, relation (1d) is preferable to (1c) but, even if one employed T/P data for the surface values K(0), the z-dependence K(z) would still be needed. On the other hand, in spite of lack of observational support, relations (1c) depend only on the resolved fields and that may explain their wide usage. The model we present below predicts the following result:

$$\kappa_M(z) = f(\overline{u},K)\ell K(z)^{1/2}, \quad \ell = \min(r_d, L_R) \tag{1e}$$

When the eddy kinetic energy K is larger than that of the mean flow (see Fig.7 of Sharffenberg and Stammer, 2010), the function $f(\overline{u},K)$ given in Eq.(15b) is unity[4].

**d)** *model assessment*. The ML mesoscale parameterization in terms of the resolved fields will be presented in terms of following variables ($\tau$ stands for an arbitrary tracer):

  Vertical-horizontal fluxes, $F_v(\tau)$, $\mathbf{F}_H(\tau)$
  Stream function-Residual flux, $\Psi$, $\mathbf{F}_r(\tau)$
    Eddy Kinetic Energy K(z)                          (2a)

Before the M parameterization is used in coarse resolution OGCMs, the following internal assessments of the model results were made. The predicted K(z) profiles are compared with WOCE data (see Figs.3), the surface value $\kappa_M(0)$ is compared with relation (1d) from Global Drifter Program/Surface Velocity Program; maps of the eddy surface energy K(0) are compared with the T/P-Jason-1data (see Figs.2). The ACC values of both K(z) and $\kappa_M(z)$ are compared with altimetry data in Figs.7.

---

[3]Gent and McWilliams (1990), Danabasoglu and Marshall (2007), Bryan et al. (1999) and Visbeck et al. (1997); the fourth is by Karsten and Marshall (2002); the last one is due to Eden et al. (2009); $\ell = \min(r_d, L_R)$ between the Rossby radius and the Rhines scale; $\mathbf{s} = -N^{-2}\nabla_H \overline{b}$ is the slope of the isopycnals, $N^2 = -g\rho_0^{-1}\rho_z$ and Ri=(f/N)$^2$s$^{-2}$ is the geostrophic Richardson number.

[4] The length scale $\ell$ is plotted in Fig.1a.

**e)** *structure of the paper*. In sec. 2 we present the derivation of the horizontal and vertical fluxes; the goal is to stress the different treatments of the non-linear terms in the deep ocean and the ML. In sec. 3, we present the forms of the vertical-horizontal fluxes for an arbitrary tracer and for the particular case of buoyancy. In sec. 3d we assess the surface eddy kinetic energy vs. T/P data; in sec. 3e we assess the model predictions of the z-profile of K(z) vs. the WOCE data, in sec. 3f we assess the mesoscale diffusivity vs. the results from the data discussed before; in sec.4 we discuss the ML re-stratification effects of mesoscales; in sec.5 we derive the analytic form of the dynamical tapering function T(z) that appears in (1b); in sec.6, we discuss the existence of a new eddy induced velocity; in sec.7 we discuss some model results for the ACC and in sec.8 we present some Conclusions.

## 2. How to model $\mathbf{F}_H, F_v$

To construct the 3D fluxes of an arbitrary tracer[5]:

$$\mathbf{F}(\tau)=\overline{\mathbf{U}'\tau'}, \quad \mathbf{F}_H(\tau)=\overline{\mathbf{u}'\tau'}, \quad F_v(\tau)=\overline{w'\tau'} \tag{3a}$$

one must first solve the dynamic equations of the mesoscale velocity and tracer fields and then take the ensemble averages of their products. At the ocean surface, the fluxes must satisfy the boundary conditions $F_v(0)=0, \mathbf{F}_H(0)\neq 0$.

In CD5,6 we discussed how to parameterize mesoscale fluxes for the case of the deep ocean. Here, we need to stress the key difference in the treatment of deep ocean mesoscales vs. ML mesoscales of interest here. The key difference is how one models the non-linear interactions represented by a turbulent viscosity and a turbulent diffusivity for the velocity and tracer fields respectively. Since in the deep ocean the flow occurs along isopycnals, what is conserved is the *potential vorticity* rather than the *relative vorticity:* enstrophy cascade (which requires the conservation of the latter) cannot occur and only an inverse energy cascade ensues which is commonly represented by a *negative* turbulent viscosity. By contrast, in the ML, the flow occurs on horizontal rather than isopycnal surfaces (Killworth, 2005). Since the former conserve *relative vorticity*, enstrophy cascade is possible which is represented by a *positive* turbulent viscosity.

**a) tracer field**

---

[5] $\mathbf{U}=(\mathbf{u},w)$ and $\mathbf{u}', w', \tau'$ are mesoscale fields representing velocity and an arbitrary tracer; an overbar denotes ensemble averages.

We begin by considering the dynamic equation for an arbitrary tracer $\tau$:

$$\partial_t \tau + \partial_i(\tau u_i) = \partial_z(K_v \partial_z \tau) + S_\tau \tag{3b}$$

The first term on the rhs is the small scale turbulent mixing with a vertical diffusivity $K_v$ and $S_\tau$ represents sources-sinks. Following standard procedure, one splits all the fields into average and fluctuating components, $\tau = \overline{\tau} + \tau'$ and $u_i = \overline{U}_i + u'_i$ and proceeds as follows. After substituting the latter into the $\tau$-equation (3b) and using the incompressibility condition, one averages the result and obtains the equation for the mean tracer. Subtracting the latter from (3b), a set of purely algebraic steps lead to the following equation for the $\tau'$ field:

$$\partial_t \tau' + \overline{\mathbf{U}} \cdot \nabla \tau' + \mathbf{U}' \cdot \nabla \overline{\tau} + Q_H^\tau + Q_V^\tau = \partial_z(K_v \partial_z \tau') \tag{4a}$$

where the Q's represent the non-linear terms defined as follows:

$$Q_H^\tau \equiv \mathbf{u}' \cdot \nabla_H \tau' - \overline{\mathbf{u}' \cdot \nabla_H \tau'}, \qquad Q_V^\tau \equiv w' \tau'_z - \overline{w' \tau'_z} \tag{4b}$$

No closure has yet been used for the Q's. Next, K5 suggested that because of the strong mixing in the ML, one can use the approximations $\overline{\tau}_z = 0$, $\tau'_z = 0$, which simplify Eq.(4a) to:

$$\partial_t \tau' + \overline{\mathbf{u}} \cdot \nabla_H \tau' + \mathbf{u}' \cdot \nabla_H \overline{\tau} + Q_H^\tau = 0 \tag{5}$$

Next, one Fourier transforms equation (5) in space and time. Following K5, we keep the same notation $\mathbf{u}', \tau'$ for the mesoscale fields in the k-$\omega$ space and assume that the mean fields $\overline{\mathbf{u}}$ and $\nabla_H \overline{\tau}$ are constant in time and horizontal coordinates. Under these conditions, the double Fourier transform of (5) reduces to the substitutions $\nabla_H \to i\mathbf{k}$, $\partial_t \to -i\omega$. Thus, we obtain:

$$i(\mathbf{k} \cdot \overline{\mathbf{u}} - \omega)\tau' + \mathbf{u}' \cdot \nabla_H \overline{\tau} + Q_H^\tau = 0 \tag{6}$$

To proceed, we need to model the non-linear term $Q_H^\tau$ without which, Eq.(6) becomes Eq.(2) of K5. To model them, we adopt the results of previous work (Canuto and Dubovikov, 1996-1999) in which a turbulence model was constructed and assessed against about 80 statistics for different turbulent flows[6]. Though the full expression of $Q_H^\tau$ is rather complex, simplifications can be obtained if one considers the region around the wave number $k_0$ where the eddy energy spectrum E(k) has its maximum. Eqs.(4e) and (8a,b) of CD5 yield the following results:

---

[6] Examples: Kolmogorov's law and constant; Batchelor's law and constant; spectra for shear driven flows; buoyancy driven flows, Nusselt number vs. Ra, 2D turbulence spectra, rotation, etc.

$$Q_H^\tau(\mathbf{k},\omega) = \chi\tau'(\mathbf{k},\omega), \quad \chi = k_0 K^{1/2}, \quad K = \frac{1}{2}\overline{|\mathbf{u}'|^2} \qquad (7)$$

where $\mathbf{u}'$ is in physical space and $\ell = k_0^{-1}$ is the mesoscale length scale discussed below[7]. To put relations (6,7) in the wider perspective of turbulence modeling, we recall that Eqs.(6), (7) represent what is known as the *stochastic Langevin equations* that were widely used in turbulence studies (Lesieur, 1990; McComb, 1992; Kraichnan, 1971, 1975; Leith, 1971; Herring and Kraichnan, 1971; Chasnov, 1991). The key advantage of the Langevin equation is that while the Navier-Stokes equations are non-linear in the velocity field with constant coefficients, the Langevin equations are the reverse, they are linear in the fluctuating fields while the coefficients in front, for example the inverse dynamical time scale $\chi$, are highly non-linear and must be modeled. The greatest advantage of the Langevin-type equations is that they allow one to compute analytically the required fluxes, whereas the highly non-linear original equation (4b) would not allow such an analytic treatment. The goal is therefore that of finding a model for the non-linear terms (4b) that leads to a Langevin equation whose correlation functions are sufficiently close to those of the original equations (6) and (4b). This is the closure problem that was discussed in detail in Canuto and Dubovikov (1997).

From the physical point of view, the first relation (7) has a simple physical interpretation within the mixing length approach: $\tilde{\chi}^{-1}$ plays the role of a dynamical time scale while the second and third relations contain the characteristic length scale and velocity. With these premises, the form of the tracer field is then given by:

$$\tau' = -\frac{\mathbf{u}' \cdot \nabla_H \overline{\overline{\tau}}}{\chi + i(\mathbf{k}\cdot\mathbf{u} - \omega)} \qquad (8)$$

Clearly, we still need an expression for the frequency $\omega$ or, more precisely, the dispersion relation which, as first discussed in K5, is obtained by solving the 1D eigenvalue problem to which the equations for the 2D mesoscale fields and thickness $h=\partial z/\partial\rho$ reduce. Since the algebraic steps to derive the eigenvalue problem were discussed in detail in K5 for the linearized case and in CD5 for the case with non-linearities, we limit ourselves to citing the final dispersion relation $\omega(\mathbf{k})=\mathbf{k}\cdot\mathbf{u}_d$ which changes Eq.(8) to

$$\tau' = -\frac{\mathbf{u}' \cdot \nabla_H \overline{\tau}}{\chi + i\mathbf{k} \cdot \mathbf{u}} \qquad \mathbf{u} = \overline{\mathbf{u}} - \mathbf{u}_d \qquad (9)$$

---

[7] replacing the kinematic $\nu$ with the turbulent one, the inverse dynamical time scale is given by $k^2 \nu_t(k) = \tilde{\nu}(k) \sim \ell^{-2}\nu_t(k) \sim \ell^{-1}K^{1/2}$, $\ell = k_0^{-1}$, from which the second relation (7) follows in the case of a unit turbulent Prandtl number $\nu_t/\chi_t = \sigma_t$. In the derivation, we have used the general "closure" (Canuto and Dubovikov, 1997) $\nu_t(k)^2 = \nu^2 + \frac{2}{5}\int_k^\infty p^{-2}E(p)dp$ which, for large Re, yields $\sim \ell^2 K$.

The dispersion relation $\omega(\mathbf{k}) = \mathbf{k} \cdot \mathbf{u}_d$ has the following physical interpretation. It represents the Doppler transformation of the frequency provided that in a system of coordinates moving with velocity $\mathbf{u}_d$, the mesoscale flow is stationary, i.e., with $\omega = 0$; this implies that $\mathbf{u}_d$ represents an *eddy drift velocity* (Henson and Thomas, 2008). It is an observed fact (Richardson, 1993) that "mesoscale eddies are water-mass anomalies with nearly circular flow around their centers which move through the background water at speeds and directions inconsistent with background flow". Stated differently, mesoscales move with a drift velocity $\mathbf{u}_d$ that in general does not coincide with the velocity $\overline{\mathbf{u}}$ of the mean flow. In CD5 we derived the following form of the drift velocity for the case of an adiabatic ocean:

$$\mathbf{u}_d^0 = <\overline{\mathbf{u}}> - \frac{1}{2} fr_d^2 \mathbf{e}_z \times <\partial_z \mathbf{s}> + \frac{1}{2} r_d^2 \mathbf{e}_z \times \boldsymbol{\beta} \qquad (10a)$$

The presence of a ML adds a contribution $\delta \mathbf{u}_d$ given by:

$$\mathbf{u}_d = \mathbf{u}_d^0 + \delta u_d \mathbf{e}_z \times \mathbf{s}(-h_*), \quad \delta u_d = -\frac{1}{2} fr_d^2 R, \quad R \equiv \frac{K(-h_*)^{1/2}}{\int_{-H}^{-h_*} K(z)^{1/2} dz} \qquad (10b)$$

where the depth $z = -h_*$ is discussed after Eq.(41a), f is the Coriolis parameter, $\boldsymbol{\beta} = \nabla f$, $\mathbf{e}_z$ is the vertical unit vector, H is the depth of the ocean and the isopycnal slope **s** is defined in footnone #3. The average $<\bullet>$ in (10a) is performed over the adiabatic part of the ocean:

$$<\bullet> \equiv \int_{-H}^{-h_*} \bullet K^{1/2}(z) dz / \int_{-H}^{-h_*} K^{1/2}(z) dz \qquad (11)$$

It must be stressed that $<\bullet>$ in (11) represents an average with a weighing factor $K^{1/2}$. Since the integration goes from the bottom of the ocean (-H) to all the way to the bottom of the ML, Eq.(11) represents a *non-local effect* of the deep ocean on the ML, or a link between the baroclinic instabilities in the ML and a barotropic variable such as $\mathbf{u}_d$. Next, we discuss the determination of the fluctuating velocity field $w'$.

**b) mesoscale velocity field**

Because of the relation $w'_z = -i k u_a$, the vertical velocity $w'$ is contributed only by the a-geostrophic component of the eddy velocity field which we must relate to the geostrophic component which is the largest of the two, $u_a \ll u_g$, and which therefore gives the largest contribution to the eddy kinetic energy, $K \approx u_g^2$. The relation between $u_a$ and $u_g$ derived in CD5, Eq.(10a), entails only algebraic manipulations of the dynamic equations for the 2D velocity field and need not be repeated here. What needs to be discussed instead is the closure of the non-linear interactions. In analogy with the tracer field (7), in the velocity case we have:

$$\mathbf{Q}(\mathbf{k},\omega) = \nu \mathbf{u}'(\mathbf{k},\omega), \quad \nu = \chi, \quad K = \frac{1}{2}\overline{|\mathbf{u}'|^2} \tag{12}$$

where we have taken a turbulent Prandtl number of unity. *The physical difference between the adiabatic deep ocean and the diabatic mixed layer is represented by the different signs with which the dynamical viscosity $\nu$ enters the equations.* In the deep adiabatic ocean, an enstrophy cascade cannot occur since the non-linear interactions do not conserve enstrophy (CD5). For that reason, there is a well-known inverse kinetic energy cascade represented by a turbulent viscosity entering the equations with a minus sign. On the other hand, in the diabatic ML, enstrophy is conserved by the non-linear interactions and this allows an enstrophy cascade which results in a *positive turbulent viscosity.* As a result, the turbulent viscosity in Eq.(10a) of CD5 enters with a negative sign and we have:

$$w'_z = -ik u_a = kf^{-1}[\mathbf{k} \cdot (\tilde{\mathbf{u}} + \mathbf{c}_R) - i\tilde{\chi}]u_g \tag{13}$$

which, once integrated over z, gives:

$$w' = zkf^{-1}[\mathbf{k} \cdot (\mathbf{u} + \mathbf{c}_R) - i\tilde{\chi}]u_g, \quad z\mathbf{u}(z) = \int_0^z \mathbf{u}(z')dz' \tag{14}$$

## 3. Vertical and Horizontal mesoscale fluxes

Once eqs.(9) and (14) are combined as discussed in Appendices A, B, we obtain:

a) horizontal flux: arbitrary tracer

We multiply relation (9) by $\mathbf{u}'$ and integrate over all k's substituting $\overline{|u_g|^2}$ with the spectrum 2E(k) and then integrate over all wave numbers k, a procedure that reduces to the substitution $2E(k) \to K$. The result is as follows:

$$\mathbf{F}_H(\tau) = -\kappa_M \nabla_H \overline{\tau} \tag{15a}$$

where the mesoscale diffusivity $\kappa_M$ is given by:

$$\kappa_M(z) = f(\bar{u},K)\ell K^{1/2}(z) \qquad f(\bar{u},K) = (1+\frac{3}{4K}|\mathbf{u}|^2)^{-1} \tag{15b}$$

where $\mathbf{u}(z)$ is defined in the second of (9).

**b) vertical flux: arbitrary tracer**

We begin by considering the z-derivative of the vertical flux:

$$\partial_z F_v = \overline{w'\partial_z\tau'} + \overline{\tau'\partial_z w'} \tag{16a}$$

whose second term can be rewritten as follows:

$$\overline{\tau'\partial_z w'} = -\overline{\tau'\nabla_H \cdot \mathbf{u}'} = -\overline{\tau'\nabla_H \cdot \mathbf{u}_a} \tag{16b}$$

where the a-geostrophic velocity $u_a$ is given by (13) in terms of the geostrophic component $u_g$. With the use of relations (9) and (14), a series of algebraic manipulation presented in Appendix B leads to the following result:

$$\frac{\partial F_v(\tau)}{\partial z} = \mathbf{u}_* \bullet \nabla_H \bar{\tau} \tag{17}$$

$$\mathbf{u}_* = \kappa_M \mathbf{F}(z) \tag{18}$$

$$\frac{1}{2}\mathrm{fr}_d^2 \mathbf{F}(z) = \mathbf{e}_z \mathbf{x}(\mathbf{u} + \frac{1}{2}z\partial_z \mathbf{u}) \tag{19}$$

Relations (19-20) exhibit an interesting physical feature. Since mesoscale eddies move with their own drift velocity $\mathbf{u}_d$, the effective mean velocity is the one in the frame co-moving with the mesoscales and that is the reason why the mean velocity enters as $\mathbf{u} = \bar{\mathbf{u}} - \mathbf{u}_d$. Integrating (17) over z, we obtain the vertical flux:

$$F_v(\tau) = -\boldsymbol{\kappa} \cdot \nabla_H \bar{\tau} \tag{20}$$

where the vector $\boldsymbol{\kappa}$ with the dimensions of diffusivity, is given by:

$$\boldsymbol{\kappa} = \kappa_M z \mathbf{F}(z) \tag{21}$$

$$\mathrm{fr}_d^2 \mathbf{F}(z) = (\mathbf{u}+\mathbf{u})\times \mathbf{e}_z, \quad z\mathbf{u}(z) \equiv \int_0^z dz'\mathbf{u}(z') \qquad (22,23)$$

It may be useful to note that at the bottom of the ML, z=-h, we have $F_v(-h) \neq 0$ since mesoscales exist throughout the entire water column.

**c) the case of a barotropic velocity field u(z)=u_b=constant.**

In this case we have $\mathbf{u} = \mathbf{u} = \mathbf{u}_b - \mathbf{u}_d$. A short algebra then yields the following result for the vertical tracer flux:

$$F_v(\tau) = z\kappa_M \mathbf{A}\cdot \nabla_H \bar{\tau}, \qquad \mathbf{A} \equiv R\mathbf{s}(-h_*) + <\partial_z \mathbf{s}> -\mathbf{\beta}f^{-1} \qquad (24)$$

which, interestingly, is independent of the barotropic velocity $\mathbf{u}_b$ which cancels out in the final result.

**d) vertical flux: buoyancy case**

For buoyancy, we take $\tau$=b in Eq.(20) and recast the result in the form:

$$F_v(b) = \kappa_v N^2 > 0, \qquad \kappa_v = \kappa_M z\mathbf{F}(z)\cdot \mathbf{s} \qquad (25)$$

where $\kappa_v$ plays the role of a *vertical mesoscale diffusivity* $\kappa_v$. The vertical buoyancy flux is plotted in Fig.1b for four locations, California Current, Gulf Stream, ACC and Labrador Sea. To illustrate the magnitude of the buoyancy flux (25), we use the values:

$\kappa_M = 10^3 \mathrm{m^2 s^{-1}}, z \approx h = 50\mathrm{m}, f=10^{-4}\mathrm{s^{-1}}, N^2=10^{-6}\mathrm{s^{-2}}, r_d=10\mathrm{km}, s=10^{-2}, \bar{u}=0.1\mathrm{ms^{-1}}, \rho=10^3 \mathrm{kgm^{-3}}$

$$c_p = 4.2\times 10^3 \mathrm{J\,kg^{-1}K^{-1}}, \; \alpha_T=10^{-4}\mathrm{K^{-1}}, \; g=10\mathrm{ms^{-2}}, \; 10^{-8}\rho c_p/g\alpha_T = 42\mathrm{Wm^{-2}(s^3 m^{-2})} \qquad (26)$$

Eq.(25) then predicts that in the middle of the ML the buoyancy and heat fluxes have values of the order of:

$$F_v(b) \approx (2\text{-}4)\times 10^{-8} \mathrm{m^2 s^{-3}}, \qquad \frac{c_p \rho}{g\alpha_T} F_v(b) \approx 100 \mathrm{Wm^{-2}} \qquad (27)$$

Contrary to the *negative* buoyancy flux due to small scale turbulence:

$$F_v^{ss} = -K_v N^2 < 0 \qquad (28)$$

the mesoscale buoyancy vertical flux (25) is *positive.* With a typical ML vertical diffusivity of $K_v=10^{-2}m^2s^{-1}$, the two fluxes turn out to be of the same order of magnitude thus possibly leading to a large cancellation.

**e) surface eddy kinetic energy, EKE**

Before we present the model results for the eddy surface kinetic energy, it is convenient to put the problem into perspective by recalling that there have been several attempts to reproduce the eddy kinetic energy with mixed results. ACC studies with the FRAM (Treguier, 1992; Stevens and Killworth, 1992; Ivchenko et al., 1996, 1997; Best et al., 1999) and $1/10^0$ resolution global eddy resolving codes (Maltrud et al., 1998), yielded values that were too low compared to the data, a shortcoming that most studies suggest it required higher resolution. To obtain the EKE, we begin by recalling the model independent dynamic equation which reads[8]:

$$\frac{DK}{Dt} + \nabla \cdot (\mathbf{M} + \overline{p'\mathbf{U'}}) = F_v(b) + \bar{S} - \varepsilon \qquad (29a)$$

where $\mathbf{M} = \overline{K\mathbf{U'}}$ is the flux of K and $\varepsilon > 0$ is the rate of dissipation of K by irreversible processes. The first term on the rhs of (29a), the mesoscale vertical flux $F_v(b)$, acts as a source of K from eddy potential energy. It may be useful to recall that such a term, with the opposite sign, appears in the equation for the eddy potential energy W:

$$\partial_t W + \nabla \cdot (\overline{\mathbf{U}}W + \overline{\mathbf{U'}W}) = -F_V(b) - N^{-2}\mathbf{F}_H(b) \cdot \nabla_H \bar{b} - \varepsilon_W \qquad (29b)$$

where it corresponds to the conversion of W into K while the second term in the rhs represents the exchange between the mean potential energy and W. The pressure terms in (29a) must be discussed in some detail. Its horizontal component $\nabla_H \cdot \overline{p'\mathbf{u'}}$ vanishes after averaging over sufficiently large region; in addition, it is of the order $10^{-9} m^2 s^{-3}$ and, therefore, does not exceed $\bar{S}$. As for the vertical term $\partial_z \overline{w'p'}$, though its average over the whole depth yields zero, since we are interested in its average over the ML, it should be considered a source of K. We begin by noting that $F_V(b) = \overline{w'b'}$ is the first term of the equality:

$$\partial_z \overline{w'p'} = \overline{w'\partial_z p'} + \overline{p'w'_z} = F_V + \overline{p'w'_z} \tag{29c}$$

and that it cancels the $F_V$ in the rhs of (29a) which then becomes (in its simplest form):

$$\partial_t K = -\overline{p'w'_z} - \varepsilon \tag{29d}$$

---

[8] $\bar{S} = -R_{ij}\bar{u}_{i,j}$, $R_{ij} = \overline{u'_i u'_j}$ (Reynolds Stresses) (Ivchenko et al., 1997; Best et al., 1999). If $\bar{S}$ (called $T_4$ in Boning and Budich, 1992), *is positive*, it represents a source of K from the mean kinetic energy (interpreted as a *barotropic instability*). In Bryden (1982), $\bar{S}$ is defined as the "transfer of mean kinetic energy to eddy kinetic energy". For the Gulf Stream, Bryden's measurements yield $\bar{S} = -(1.5 \pm 0.9) \cdot 10^{-9} m^{-2}s^{-3}$ which means that eddies lose energy to the mean flow. On the other hand, in the ACC (Treguier, 1992; Ivchenko et al., 1997; Best et al., 1999), $\bar{S} > 0$ and eddies gain energy from the mean flow. In the ACC, $\bar{S}$ contributes only 8% of the eddy kinetic energy production (Ivchenko et al., 1997). Eden and Greatbach (2008) mesoscale resolving simulations show both $F_v$ and $\bar{S}$ (in units of $10^{-9} m^{-2}s^{-3}$) at a 200m; $F_v$ is largely positive in the North Atlantic with peak values over $10^2$ along the Gulf Stream while $\bar{S}$ is mostly negative. In general, $\bar{S}$ is an order of magnitude smaller than $F_v$.

In Appendix C we derive the relation:

$$-\overline{p'w'_z} = 2\lambda(z) r_d^{-1} K^{3/2} \tag{29e}$$

where $\lambda(z)=\pm 1$, the upper sign corresponds to the ML while the lower one to the ocean interior. Thus, in the ML, EKE is produced while in the ocean interior it is destroyed and transformed into EPE (eddy potential energy), in accordance with the description presented in CD5. Equation (29d) implies that $-\overline{p'w'_z}$ is the source of EKE in the ML. On the other hand, Eq.(29a) also implies that when one considers the whole ocean, the source of EKE is $F_V$. This suggests that such variables, when averaged over some region of depth D, are proportional to each other, i.e.:

$$\int_{-D}^{0} F_V dz = 2C^{-1} r_d^{-1} \int_{-D}^{0} K^{3/2} dz \tag{29f}$$

This relation can be compared with the one obtained by integrating (29c) from –H to 0, using (29e) and the boundary condition $w'(0,-H)=0$, namely:

$$(\int_{-H}^{-h_*} + \int_{-h_*}^{0}) F_V dz = 2r_d^{-1} (\int_{-h_*}^{0} K^{3/2} dz - \int_{-H}^{-h_*} K^{3/2} dz) \tag{29g}$$

While in the lhs there are two positive terms, in the rhs there is the difference of two positive contributions of approximately the same order of magnitude. Therefore, it must be true that:

$$2r_d^{-1} \int_{-h_*}^{0} K^{3/2} dz \gg \int_{-h_*}^{0} F(z) dz \tag{29h}$$

Thus, if D is close to $h_*$, we can conclude that in (29f) $C > 1$. We chose D to be the depth where $F_V(z)$ reaches its maximum value and assume that within such a depth the profile $K(z) \approx K_s$ is almost z-independent. Substituting Eqs.(20), (22) and (15b) into (29f), we obtain:

$$K_s = -\frac{1}{2} r_d^2 C <z\mathbf{F} \cdot \nabla_H \overline{b}>, \qquad <\bullet> \equiv \frac{1}{D} \int_{-D}^{0} \bullet \, dz \qquad (30a)$$

where $\mathbf{F}(z)$ is given in Eq. (23). It is worth noting that analogous relations also apply to sub-mesoscales which were studied in CD10 using the simulation data of Capet et al. (2008) with C=16. Substituting relations (22-23) and (15b) into (30a), we obtain the final expression for the surface eddy kinetic energy:

$$K_s = f^{-1} <|z|(\mathbf{u}+\mathbf{u}) \times \mathbf{e}_z \cdot \nabla_H \overline{b}> \qquad (31a)$$

$$\mathbf{u} + \mathbf{u} = z^{-1} \int_0^z \overline{\mathbf{u}}(z) dz + \overline{\mathbf{u}}(z) - 2\mathbf{u}_d \qquad (31b)$$

where we used the definitions of $\mathbf{u}, \mathbf{u}$ given in Eqs.(9) and (20). Some comments are in order: **a)** (31) does not involve a length scale, **b)** Eq.(31) emphasizes the fact that in a baroclinic regime, the eddy kinetic energy is proportional to the *slope variability* represented by the horizontal gradient of the mean buoyancy, in contrast to the eddy potential energy EPE which is proportional to the *height variability* $(\text{EPE} \sim \overline{h'^2}/N^2)$ **c)** the term $\mathbf{u}_d$ that enters $\mathbf{F}(z)$ in Eq.(23) through the second of (9), is contributed by the deep ocean since in Eq.(11) one integrates from the bottom of the ocean –H, **d)** the other terms in $\mathbf{F}(z)$ are contributed by the ocean upper layers (Ekman layer[9])

**f) assessment of the surface EKE using the T/P data**

In Fig.2a we reproduce the 3-years average surface eddy kinetic energy from the Topex-Poseidon-Jason-1 data (Scharffenberg and Stammer, 2010). In Figs.2b we show the map of $K_s$ from Eq.(31). The present model seems to be able to reproduce eddy kinetic energies of the right magnitude and geographic distribution: in fact, the results in Fig.2b exhibit structures and features quite similar to the data in Fig.2a with comparable strength in the Gulf Stream and the ACC. In Fig.2c we present $K_s$ without the contribution of the deep ocean while Fig.2d corresponds to the case with only the deep ocean contribution. While Fig.2c yields results that are excessively large everywhere, the results in case in Fig.2d are negative in the southern ocean and counterbalance the large values in Fig.2c, the net result being Fig.2c[10]. Finally, in Fig.2e we present the zonal averages of the surface kinetic energy, altimetry data and present model. The similarity of the results is reassuring. In order to further assess the comparison between the data and the model results, we computed the root mean square of the surface $K_s$:

$$K_{rms} = \left(\frac{\Sigma \alpha_i K_i^2}{\Sigma \alpha_i}\right)^{1/2} \qquad (32)$$

where $\alpha_i$ is the area of the grid cell and weights the rms. In most places the difference K(data) – K(model) is ± 50 cm²s⁻². The RMS of K for the data and our model are 312cm²s⁻² and 286cm²s⁻² respectively, proving the model's ability to estimate K. The RMS difference:

$$K_{rms}(\text{data-model}) = \left(\frac{\Sigma \alpha_i [K_i(\text{data}) - K_i(\text{model})]^2}{\Sigma \alpha_i}\right)^{1/2} = 212 \text{cm}^2\text{s}^{-2} \qquad (33)$$

---

[9] To visualize the role of the two velocities, we computed the z-profile of the ratio $f|\overline{\mathbf{u}}_z|/|\nabla_H \overline{b}|$ in several regions. For the month of February, in the ACC, it becomes unity (geostrophy) from a depth of 0.4h down (h is the ML depth), in the Sea of Japan from 0.3h down, in the Gulf Stream from 0.3h down, in the Labrador Sea from 0.5h down, showing that $\mathbf{F}_B$ is contributed by both geostrophy and Ekman flow in different regions.

The correlation (data, model) is 0.51 and the covariance (data, model) is 43325 (cm²s⁻²)².

**g) K(z) profile: assessment using WOCE data**

The model prediction of the K(z) profile derived in CD5,6, is as follows:

$$K(z)/K_s = \Gamma(z), \quad \Gamma(z) = (a_0^2 + |B_1(z)|^2)(1 + a_0^2)^{-1}, \quad a_0^2 \approx \overline{K}_{ML}/K_s \qquad (34)$$

where $\overline{K}_{ML}$ is the mean kinetic energy $\overline{K}$ averaged over the ML. To compute the first baroclinic mode $B_1(z)$, one must solve the eigenvalue equation:

$$\partial_z(N^{-2}\partial_z B_1) + (r_d f)^{-2} B_1(z) = 0 \qquad (35)$$

with the boundary conditions $B_1(-h)=1, \partial_z B_1 = 0$ at z= -h, -H (defined earlier). Eq.(34) is in accordance with Wunsch (1997) who showed that in the vicinity of the thermocline, the mesoscale velocity field is contributed mostly by the first baroclinic mode $B_1(z)$, and thus, to first order, $\Gamma(z) \approx |B_1(z)|^2$ while below the thermocline, the contribution of the barotropic modes becomes important. In **Figs.3a,b,** we present the K(z) profiles given by relations (34-35) in different oceans basins. The data are from WOCE (WOCE Data Products Committee. 2002. WOCE Global Data, Version 3.0, WOCE International Project Office, WOCE Report No.

---

[10] The mean fields (mean velocity, T, S ,density, $N^2$, mixed layer depth, slopes **s**), were obtained from the GISS atmosphere-ocean coupled model discussed in detail in Hansen et al. (2007). The vertical mixing model is the one recently presented in (Canuto et al., 2010, 2011).

180/02, Southampton, UK). The same data also provide the mean of the time series used to compute $\overline{K}_{ML}$ in Eq.(34). The comparison is satisfactory.

**h) mesoscale diffusivity $\kappa_M(z)$**

As Eq.(1c) shows, several heuristic expressions have been proposed to determine $\kappa_M(z)$ e.g., by Visbeck et al., (1997), Stammer (1998), Zhurbas and Oh (2003, 2004), Marshall et al. (2006), Rupolo (2007), Sallee' et al. (2008a,b; 2010). The present model predicts relations (15b). Using the model for the surface eddy kinetic energy given by relations (31), in **Figs.4a** we plot the surface values of the mesoscale diffusivity. Using relations (34-35), in **Fig.4b** we show the z-profile of $\kappa_M(z)$ normalized to the surface value. It is interesting to note how the curves collapse into an "almost universal" function even though they correspond to quite different geographical locations. In **Fig.4c** we present maps of the diffusivities at different depths where the values are limited to narrower bands than near the surface. Further model results for the ACC are presented in **Figs.7.**

## 4. Mixed layer re-stratification

The time evolution of the stratification is governed by the equation:

$$\frac{\partial N^2}{\partial t} = -\frac{\partial^2 F_v}{\partial z^2} - \frac{\partial^2 F_v^{ss}}{\partial z^2} \dots \quad (36)$$

where, for the present purposes, we have kept only the vertical mesoscale and the small scale fluxes (denoted by ss) which always de-stratify the ML. Whether mesoscales de-stratify or re-stratify the ML depends on the sign of the second z-derivative of the vertical flux $F_v(b)$. In Eqs.(20-23), the main contribution to (36) is determined only by the large scale velocity in the ML since the contribution of the eddy drift velocity $\mathbf{u}_d$ entering in $\mathbf{u}$ is proportional z which does not contribute to $\partial_{zz} F_V$. Below the Ekman layer, the mean velocity is geostrophic and thus $\partial_{zz} F_V \sim -|\nabla_H \overline{b}|^2 < 0$ which implies ML re-stratification. The Ekman layer does not change the conclusion that $\partial_{zz} F_V < 0$. The mesoscale re-stratification of the ML predicted by the present model is in accordance with previous studies (Nurser et al., 2000; Oschlies, 2002; Hosegood et al., 2008).

## 5. The case of Buoyancy

Since in the ocean interior, mesoscale fluxes are traditionally parameterized in terms of a stream function and residual flux $\Psi$, $\mathbf{F}_r$ of the residual mean theory RMT, to match the ML with the ocean interior, we need to express the vertical-horizontal fluxes obtained above, Eqs.(15a) and (20), in terms of $(\Psi, F_r)$. The problem was treated elsewhere (Canuto and Dubovikov, 2011, CD11) where it was shown that, due to the diabatic nature of the ML, the traditional definition of $\Psi$ in terms of mesoscale fields (Andrew and McIntyre, 1976; Plumb and Ferrari, 2005; Ferrari et al., 2008) does not satisfy the boundary condition $\Psi(0) = 0$ as required in order for the vertical eddy induced velocity to vanish at the surface. In order to satisfy the $\Psi(0) = 0$ requirement, the traditional $(\Psi, F_r)$ must be substituted by a modified stream function and residual flux denoted by $\Psi, F_r$. Specifically, for a horizontal buoyancy flux $F_H$ of the form (15a), the stream function is given by the following expression:

$$\Psi = -(sN)^{-2} F_v(b) \mathbf{s} \times \mathbf{e}_z \tag{37}$$

Substituting (20) for the buoyancy field, we derive the relations[11]:

$$\Psi(z) = \mathbf{Y}(z) \times \mathbf{e}_z, \quad \mathbf{Y}(z) = -\kappa^\ell = -z\kappa_M \frac{\mathbf{F} \cdot \nabla_H \bar{b}}{|\nabla_H \bar{b}|^2} \nabla_H \bar{b} \tag{38}$$

or, alternatively:

$$\mathbf{Y}(z) = T(z)\, \mathbf{Y}(z), \quad T(z) = -z \frac{N_*^2}{|\nabla_H \bar{b}|^2} \mathbf{F} \cdot \nabla_H \bar{b},$$

$$\mathbf{Y}(z) = \kappa_M \frac{\nabla_H \bar{b}(z)}{N_*^2} = -\kappa_M \mathbf{s}_* \frac{\nabla_H \bar{b}(z)}{\nabla_H \bar{b}(z)_*} \tag{39}$$

where $N_*^2 \equiv N^2(z=-h_*)$, $\mathbf{s}_* \equiv \mathbf{s}(z=-h_*)$ where $h_*$ is defined below in Eq.(41a). The eddy induced velocity follows from the relation:

$$\mathbf{U}^+ = \nabla \times \Psi = \nabla \times [\mathbf{Y}(z) \times \mathbf{e}_z]$$

$$\mathbf{u}^+ = \frac{\partial \mathbf{Y}}{\partial z} = -\frac{\partial \kappa^\ell}{\partial z} = \frac{\partial}{\partial z}[T(z)\, \mathbf{Y}(z)] \quad, \quad w^+ = -\nabla_H \cdot \mathbf{Y} \tag{40a}$$

The function T(z) in the first of (39) is no longer arbitrary, it is given by the model and satisfies the conditions $T(0) = 0$ which implies that $\Psi(0) = 0$; T(z) increases with depth and reaches unity at the depth $h_*$ where the diapycnal flux vanishes:

$$F_d \equiv F_V + N^{-2} \mathbf{F}_H \cdot \nabla_H \bar{b} = F_V - \mathbf{F}_H \cdot \mathbf{s} = 0 \tag{40b}$$

---

[11] $\ell$ is the longitudinal (with respect to the direction of **s**) component of a vector $\mathbf{V}$, i.e., $\mathbf{V}^\ell = s^{-2}(\mathbf{V} \cdot \mathbf{s})\mathbf{s}$

which leads to:
$$T(-h_*) = 1 \tag{41a}$$

Thus, the depth $-z=h_*$ represents the boundary between the ML and the deep ocean.

A physical interpretation of the above results may be useful. If we use the first of (38) and the first and third relations in (39), we can recast the new stream function in the form:

$$\mathbf{\Psi}(z) = -\kappa_M T(z) \mathbf{s}_* \times \mathbf{e}_z \left( \frac{\nabla_H \overline{b}(z)}{\nabla_H \overline{b}(z)_*} \right) \tag{41b}$$

If we compare (41b) with the GM model (1b), we reach the following conclusions: first, while in (1b) the tapering function $T(z)$ is not known, in the present model it is given by the second relation (39) in terms of the resolved fields and it is thus calculable; second, since in the practical applications of (1b) the slope $\mathbf{s}$ is taken to be $\mathbf{s}_*$ at the bottom of the ML, the new feature in (41b) is the last term representing the ratio of the buoyancy gradient at any z within the ML to its bottom value.

Next, we must construct the ML residual buoyancy flux which is computed from its definition (CD11):

$$\mathbf{F}_r(b) = \mathbf{F}(b) - \mathbf{\Psi} \times \nabla \overline{b} \tag{42a}$$

Using (15), (20) and (38)-(39), relation (42a) becomes:

$$\mathbf{F}_r = -\kappa_r \nabla_H \overline{b}, \quad \kappa_r = \kappa_M [1 - pT(z)], \quad p \equiv \frac{N^2(z)}{N^2_*} \tag{42b}$$

It is clear that $T(z)$ also satisfies (41a) and therefore the residual mesoscale diffusivity $\kappa_r$ vanishes at $h_*$. Relations (42) show that the z-profile of $\kappa_r$ and the tapering function $T(z)$ are relate to one another while in the default tapering scheme they are independent. The nice feature of relations (38) and (42b) is that at $z=-h_*$ they smoothly match the deep ocean GM parameterizations represented by:

$$\mathbf{\Psi}(-h_*) = \mathbf{\Psi}_{GM}(-h_*), \quad \mathbf{F}_r(-h_*) \approx 0 \tag{43a}$$

In other words, at the boundary of the ML with the adiabatic ocean interior, the new stream-function naturally matches the GM parameterization in the adiabatic ocean with no additional phenomenological adjustment[12] while the residual flux vanishes.

---

[12] the GM parameterization is not the only one able to describe the ocean interior. If we employ the dynamical mesoscale model presented in CD5,6, we impose the same boundary condition on the ocean interior stream

Several considerations are in order. Results (39) and (42b) are similar to those of the heuristic tapering schemes which contain three unknown variables, the tapering function T(z), the residual diffusivity $\kappa_r(z)$ and the depth $h_*$ where tapering begins (Griffies et al., 2005; Gnanadesikan et al., 2007). The tapering function T(z) is usually assumed to be a linear function of z and unlike the present model, where $\mathbf{Y}(z)$ is a function of z, in such models it is taken to be z-independent and $h_*$, $\mathbf{Y}(-h_*) \to \mathbf{Y}_{GM} = -(\kappa_M \mathbf{s})_*$, see, for example, Eq.A.13 of Marshall et al. (2006) whose function $\mu(z)$ plays the role of T(z). In all previous models, the depth $h_*$ was chosen from the condition that below it, $|\mathbf{s}| \leq s_{max}$, which is equivalent to selecting an $s_{max}$, a choice that turned out to be difficult and consequential since numerical simulations showed that the results were sensitive to choices such as $s_{max}$ = 1/500, 1/100, (Gnanadesikan et al., 2007; Ferrari et al., 2008). By contrast, in the present model, the large scale fields uniquely determine T(z), $\kappa_r(z)$ and $h_*$. The tapering function T(z) is plotted in **Fig.5** at two different locations at a time corresponding to the deepest mixed layer.

In conclusion, in the buoyancy case which we studied so as to compare our model results with those of previous parameterizations, the residual flux is purely diffusive and there is only one eddy induced velocity. How these conclusions change in the case of an arbitrary tracer, which is the one of interest in OGCMs, is discussed in the next section.

## 6. The case of an arbitrary tracer

Since OGCMs do not time step buoyancy but tracers such as temperature, salinity, CO2, CFC etc, which cannot be represented by the buoyancy field, it is necessary to treat the case of an arbitrary tracer described by the dynamic equation:

$$\partial_t \bar{\tau} + \overline{\mathbf{U}} \cdot \nabla \bar{\tau} + \nabla \cdot \mathbf{F}(\tau) = -\nabla \cdot \mathbf{F}_{SM} - \partial_z F_{ss} + Q \qquad (44a)$$

---

function: $\mathbf{\Psi}(-h_*) = \mathbf{\Psi}_{GM}(-h_*)$ which allowed us to derive a correction to expression (10a) for $\mathbf{u}_d$. To this end, we use the fact that the expression for the eddy induced velocity in terms of the eddy drift velocity $\mathbf{u}_d$, Eqs.(4a,d) of CD6, $\mathbf{u}^+ = \kappa_M[\partial_z \mathbf{s} - 2(fr_d^2)^{-1}(\mathbf{u}_d - \bar{\mathbf{u}}) \times \mathbf{e}_z + f^{-1}\mathbf{\beta}]$ does not depend on boundary conditions. Using the first of (40a) and $\mathbf{Y}(-H)=0$, we obtain $\int_{-H}^{-h_*} \mathbf{u}^+(z)dz = \mathbf{Y}_{GM}(-h_*) = -\kappa_M(-h_*)\mathbf{s}_*$. Substituting the expression for $\mathbf{u}^+$ and $\kappa_M(z) = r_d K(z)^{1/2}$, we arrive at result (10), (11). It is worth noticing that, strictly speaking, we have to account also for the effect of the bottom boundary layer. Then, in the left hand side of (43c) we have to add the term $-\kappa_M(-H)\mathbf{s}(-H)$ where H is not the full ocean depth but the distance between the ocean surface and the beginning of the bottom boundary layer. It is easy to find the additional correction to (10b).

where on the rhs we have added, for completeness, the contributions from sub-mesoscales (SM), small scale turbulent mixing (ss) and sources and sinks (Q). Using relations (15a) and (20), the total mesoscale flux is given by:

$$\mathbf{F}(\tau) = \mathbf{F}_H + F_v \mathbf{e}_z = -\kappa_M \nabla_H \bar{\tau} - (\boldsymbol{\kappa} \cdot \nabla_H \bar{\tau})\mathbf{e}_z \tag{44b}$$

We recall that (20) was derived under the assumption of a small ML stratification $N^2(z)$ which, in the vicinity of $z=-h_*$, is no longer valid since $N^2(z)$ grows significantly in that region. Thus aa correction of the order of $N^2(z)/N_*^2)$ is needed so as to match the parameterization of the mesoscale tracer flux in the ocean interior. As we show below, the transverse component of the diffusivity $\boldsymbol{\kappa}$ must be changed to:

$$\boldsymbol{\kappa}^{tr} \rightarrow [1-pT(z)]\boldsymbol{\kappa}^{tr} \tag{45}$$

Near the surface, we have that $N^2(z) \ll N_*^2$ and $T(z)<1$ and one can see from (42b) that the change of $\boldsymbol{\kappa}^{tr}$ in (45) is negligible and (44b) does not change. On the other hand, at $z=-h_*$ we have $pT(z)=1$ and the transverse component $\boldsymbol{\kappa}^{tr}$ vanishes and only the longitudinal contribution $\boldsymbol{\kappa} \rightarrow \boldsymbol{\kappa}^{\ell}$ survives[13]. Taking into account (45), the tracer flux (44b) acquires the following form:

$$\mathbf{F}(\tau) = \boldsymbol{\Psi} \times \nabla \bar{\tau} + \mathbf{F}_r(\tau) \tag{46a}$$

$$\mathbf{F}_r(\tau) = \mathbf{F}_r(\tau) + \mathbf{F}_r(\tau,b) \tag{46b}$$

$$\mathbf{F}_r(\tau) = -\kappa_r \nabla_H \bar{\tau} \tag{46c}$$

$$\mathbf{F}_r(\tau,b) = -\kappa_M pT(z)(\nabla_H \bar{\tau} - \frac{\bar{\tau}_z}{N^2}\nabla_H \bar{b}) + \mathbf{Y}_{**} \cdot \nabla_H \bar{\tau}\mathbf{e}_z \tag{46d}$$

$$\mathbf{Y}_{**} = -[1-pT(z)]\boldsymbol{\kappa}^{tr} \tag{46e}$$

where the stream function $\boldsymbol{\Psi}$ and the residual diffusivity $\kappa_r$ are given in (38)-(39) and (42b). Notice that since $\boldsymbol{\kappa}^{tr} \cdot \nabla_H \bar{b} = 0$, we have $\mathbf{F}_r(b,b)=0$. To further understand the residual flux, we rewrite the last term of (46d) as follows:

$$\mathbf{Y}_{**} \cdot \nabla_H \bar{\tau}\mathbf{e}_z = \mathbf{Y}_{**} \cdot \nabla \bar{\tau}\mathbf{e}_z = \mathbf{F}^{**}_{skew}(\tau) + \mathbf{Y}_{**}\bar{\tau}_z \tag{47a}$$

---

[13] Since $\boldsymbol{\kappa}^{tr} \cdot \nabla_H \bar{b} = 0$, in the case of buoyancy the substitution (45) is not needed.

where:

$$\mathbf{F}_{skew}^{**}(\tau) = \mathbf{\Psi}_{**} \times \nabla\overline{\tau}, \quad \mathbf{\Psi}_{**} = \mathbf{Y}_{**} \times \mathbf{e}_z \quad (47b)$$

This means that we can regroup the terms in the residual flux (46b) so as to exhibit diffusive and skew components, specifically:

$$\mathbf{F}_r(\tau) = \mathbf{F}_{diff}(\tau) + \mathbf{F}_{skew}^{**}(\tau) \quad (48a)$$

$$\mathbf{F}_{diff}(\tau) = -\kappa_r \nabla_H \overline{\tau} - \kappa_M pT(z)(\nabla_H \overline{\tau} - \frac{\overline{\tau}_z}{N^2}\nabla_H \overline{b}) + \mathbf{Y}_{**}\overline{\tau}_z \quad (48b)$$

While in the buoyancy case, Eq.(42b), the residual flux is purely diffusive, the tracer residual $\mathbf{F}_r(\tau)$ given by (48a) contains not only a diffusive component but also a skew one whose divergence in the tracer equation (44a) yields an advection with the bolus velocity:

$$\mathbf{U}_{**} = \nabla \times \mathbf{\Psi}_{**} = (\mathbf{u}_{**}, w_{**}):$$

$$\mathbf{u}_{**} = \partial_z \mathbf{Y}_{**}, \quad w_{**} = \nabla_H \cdot \mathbf{Y}_{**} \quad (49)$$

Thus, the final form of the ML tracer equation reads as follows:

$$\partial_t \overline{\tau} + (\overline{\mathbf{U}} + \mathbf{U}^+ + \mathbf{U}_{**}) \cdot \nabla \overline{\tau} + \nabla \cdot \mathbf{F}_{diff}(\tau) = -\nabla \cdot \mathbf{F}_{SM} - \partial_z F_{ss} + Q \quad (50)$$

Eq.(50) contains two bolus velocities $\mathbf{U}^+$ and $\mathbf{U}_{**}$ given by Eqs. (40) and (49), the first originating from the stream function while the second, a new feature, comes from the skew component of the residual flux (48a). Notice that in order to eliminate the apparent singularity of (48b) when $N^2 \to 0$, in numerical simulations one can use the last of (42b) and substitute (48b) with the equivalent expression:

$$\mathbf{F}_{diff}(\tau) = -\kappa_r \nabla_H \overline{\tau} - \kappa_M \frac{T(z)}{N_*^2}\left[N^2(z)\nabla_H \overline{\tau} - \overline{\tau}_z \nabla_H \overline{b}\right] + \mathbf{Y}_{**}\overline{\tau}_z \quad (51a)$$

or more transparently:

$$\mathbf{F}_{diff}(\tau,b) = -\kappa_M \nabla_H \overline{\tau} - \kappa_M \mathbf{\Omega}(\tau,b)\frac{\partial \overline{\tau}}{\partial z} \quad (51b)$$

where the vector $\mathbf{\Omega}(\tau,b)$ is given by:

$$\mathbf{\Omega}(\tau,b) = z[1-pT(z)]\mathbf{F}(z) - \frac{p}{N_*^2}T(z)^2 \nabla_H \overline{b} \quad (51c)$$

We notice the following limits:

$$z=0: \quad \mathbf{F}_{\text{diff}}(\tau,b) = -\kappa_M \nabla_H \bar{\tau} \quad (51d)$$

$$z=-h_*: \quad \mathbf{F}_{\text{diff}}(\tau,b) = -\kappa_M \nabla_H \bar{\tau} + \kappa_M \frac{\nabla_H \bar{b}}{N^2}\left(\frac{\partial \bar{\tau}}{\partial z}\right), \quad \mathbf{F}_{\text{diff}}(b,b) = 0 \quad (51e)$$

We must also remark that in the buoyancy case:

$$\mathbf{U}_{**} \cdot \nabla \bar{b} = 0 \quad (52)$$

and thus the new bolus velocity $\mathbf{U}_{**}$ acts only on tracers other than buoyancy. This means that the assumption (e.g., Ferreira and Marshall, 2006) that there is only one bolus velocity $\mathbf{u}^+$ is no longer valid. The so-called "*residual velocities*" are now defined as follows:

$$\mathbf{u}_{\text{res}}(\text{old}) = \bar{\mathbf{u}} + \mathbf{u}^+, \quad \mathbf{u}_{\text{res}}(\text{new}) = \mathbf{u}_{\text{res}}(\text{old}) + \mathbf{u}_{**} \quad (53)$$

In **Fig.6a,b** we show the z-profile of the x-y components of the old and new residual velocities. In Fig.6a, the difference between old and new residual velocities is not large below say 45m but it increases in size as one approaches the surface with $u_{\text{res}}(\text{new})$ being actually smaller than $u_{\text{res}}(\text{old})$. In both cases, however, the residual velocities are positive. In the case of the y-component, going from the bottom of the ML toward the surface, one observes that $v_{\text{res}}(\text{new})$ is more negative than $v_{\text{res}}(\text{old})$; $v_{\text{res}}(\text{new})$ can reach values more up to three times as large as $v_{\text{res}}(\text{old})$. Both residual velocities decrease in magnitude toward the surface but while $v_{\text{res}}(\text{old})$ becomes positive in the last 20 m and then remains positive, $v_{\text{res}}(\text{new})$ stays negative; while at about 15m depth, $v_{\text{res}}(\text{old})$ is zero, $v_{\text{res}}(\text{new})$ is still negative and reaches its maximum negative value of about 5cms$^{-1}$. The effect of the *residual velocity* on the time variation of a mean tracer $\bar{\tau}$ can be seen by considering the two advection terms:

$$\partial_t \bar{\tau} = -u_{\text{res}} \partial_x \bar{\tau} - v_{\text{res}} \partial_y \bar{\tau} + \ldots \quad (54)$$

Since both old and new $u_{\text{res}}>0$ and approximately of the same magnitude, in the $\partial_x \bar{\tau}>0$ case, both residual velocities act as a sink of mean $\bar{\tau}$ while the opposite is true if $\partial_x \bar{\tau}<0$. In the $v_{\text{res}}$ case, when $\partial_y \bar{\tau}>0$, the more negative $v_{\text{res}}(\text{new})$ implies a *larger source* of C than $v_{\text{res}}(\text{old})$, while it entails a *larger sink* in the case of $\partial_y \bar{\tau}<0$.

Finally, consider the ML expression (46a) at the boundary $z=-h_*$ between the diabatic ML and adiabatic ocean interior. With account of (43a), Eq.(46a) becomes:

$$\mathbf{F}(-h_*) = \mathbf{\Psi}_{GM}(-h_*) \times \nabla \bar{\tau} + \mathbf{F}_r(-h_*) \quad (55a)$$

Using (46b-e) and the second of (42b) with $T(-h_*) = \overline{T}(-h_*)=1$, we obtain:

$$\mathbf{F}_r(-h_*) = -\kappa_M[\nabla_H\overline{\tau} - \overline{\tau}_z N_*^{-2})\nabla_H\overline{b}(-h_*)] \tag{55b}$$

In summary, relations (46) give the ML tracer flux; at the bottom of the ML, conditions (55a,b) hold true.

## 7. The ACC

Over the years, the ACC has been the subject of many studies (Treguier, 1992; Stevens and Killworth, 1992; Ivchenko et al., 1996, 1997; Best et al., 1999; Marshall and Radko, 2003; Marshall et al., 2006; Salle'e et al., 2008a,b, 2010, 2011). In particular, the mesoscale kinetic energy and the mesoscale diffusivities were extensively studied using numerical simulations and analytic models. In **Fig.7a** we present the surface kinetic energy from altimeter data (Scharffenberg and Stammer, 2010) and the present model (bottom figure). The magnitudes are quite similar. In **Fig.7b** we present maps of the ACC mesoscale diffusivities at different depths. The surface values shown in **Fig.7b** are about an order of magnitude lager than those of **Fig.5a** of Marshall et al.(2006) that do not exceed $\sim 10^3 m^2 s^{-1}$. The present values are closer to the data. In **Fig.7c** we present the tapering function T(z) computed at the bottom of the ML for January, July and annual average. Finally, the stream wise integral of the stream function is given by (r is the earth's radius)[14]:

$$\Psi_{\text{steamwise}} \equiv \int d\mathbf{l} \cdot \mathbf{\Psi} = r\sin\varphi(\int d\theta\cos\theta\Psi_y - \int d\theta\sin\theta\Psi_x) \tag{56a}$$

Using relations (38)-(39) we obtain:

$$\Psi_x = T(z)\, Y_y, \quad \Psi_y = -T(z)Y_x \tag{56b}$$

which we plot in **Fig.7d** ($1Sv=10^6 m^3 s^{-1}$). The results for $\Psi_x$ can be compared with the lowest curve in Fig. 7 of Marshall et al. (2006; there is an overall sign difference in the definitions). Since at the bottom of the ML the tapering function is unity, the stream function is the product of the mesoscale diffusivity with the slope of the isopycnals. The first ingredient is predicted by the present model while it is derived using heuristic arguments in Marshall et al. (2006). The second component, the isopycnal slopes, are derived from the coupled atmosphere-ocean OGCM used in this work while in Marshall et al. it is taken from the Gouretski and Janke (1998) climatology. We consider it quite encouraging that the maximum value and its location predicted by the present model are quite close to the heuristic results of Marshall et al. The difference in the magnitude and shape of the curve at latitudes less than $45^0$ must be due to the different estimates of the isopycnal slopes and it may be an instructive information on the limitation of the coupled-model used in this work.

---

[14] Consider the line element $d\mathbf{l} = \hat{\mathbf{r}}dr + \boldsymbol{\varphi}\, rd\varphi + \boldsymbol{\theta}\, r\sin\varphi d\theta$, where θ=longitude, φ=π/2-δ(latitude). Changes in longitude only correspond to: $d\mathbf{l} = \boldsymbol{\theta} r\sin\varphi d\theta$, where $\boldsymbol{\theta} = (-\sin\theta, \cos\theta, 0)$.

## 8. Conclusions

Using the solutions of the ML mesoscale dynamic equations with the non-linear terms, we constructed the horizontal-vertical mesoscale fluxes for an arbitrary tracer. The vertical buoyancy flux was used to construct the mixed layer mesoscale kinetic energy. The horizontal-vertical mesoscale fluxes for an arbitrary tracer were used to construct the stream function and residual flux of the RMT. Some of the results can be summarized as follows:
a) the vertical flux automatically vanishes at the surface avoiding the need for tapering functions,
b) the model predicts that mesoscales re-stratify the mixed layer, in agreement with eddy resolving simulation data,
c) the predicted z-profile of the eddy kinetic energy and its surface value compare well with WOCE and altimeter data T/P data,
d) the z-profile of the mesoscale diffusivity is computed and its surface value is compared with that obtained from GDP(Global Drifter Program) data for the ACC,
e) in the case of an arbitrary tracer, the residual flux contains a skew component that generates a new eddy induced velocity to be added to the one computed as the curl of the stream function,
f) thus, the common assumption that the eddy induced velocity for buoyancy can also describe an arbitrary tracer is not correct in the ML where active (e.g., buoyancy) and passive (e.g., CO2) tracers are described by different bolus velocities,
g) an analytic expression for the tapering function T(z) is derived in terms of the large scale variables; T(z) vanishes at the surface and tends to unity below the ML where the stream function smoothly connects with the deep ocean GM form.
    The next step is the implementation of the above model in a coarse resolution OGCM.

### Appendix A. Horizontal flux

The general procedure consists of computing the functions in $(\omega, \mathbf{k})$-space which, in the approximation of homogenous and stationary mean flow, have the form:

$$\overline{A'(t,\mathbf{k}')B'^{*}(t,\mathbf{k})} = \overline{A'B'^{*}}(\mathbf{k})\delta(\mathbf{k} - \mathbf{k}') \tag{A1}$$

The function $\text{Re}\overline{A'B'^*}(\mathbf{k})$ is called the density of $\overline{A'B'}$ in k-space. The spectrum of the correlation function $\overline{A'B'}$ is then given by:

$$\overline{A'B'}(k) = \int \text{Re}\overline{A'B'^*}(\mathbf{k})\delta(k-|\mathbf{k}|)d^2\mathbf{k} \tag{A2}$$

i.e., the spectrum is obtained by averaging $\text{Re}\overline{A'B'^*}(\mathbf{k})$ over the directions of $\mathbf{k}$ and multiplying the result by $\pi k$. Finally, the correlation function $\overline{A'B'}$ in physical space is obtained by integrating its spectrum. In order to compute the horizontal tracer flux $\mathbf{F}_H$ in accordance with the above strategy, we begin by computing its density in k-space:

$$\mathbf{F}_H(\mathbf{k}) = \text{Re}\overline{\mathbf{u}'\tau'^*}(\mathbf{k}) \tag{A3}$$

where, in accordance with (A1), we have:

$$\text{Re}\overline{\mathbf{u}'(t,\mathbf{k})\tau'^*(t,\mathbf{k}')} = \text{Re}\overline{\mathbf{u}'\tau'^*}(\mathbf{k})\delta(\mathbf{k}-\mathbf{k}') \tag{A4}$$

Substituting (9), we obtain:

$$\mathbf{F}_H(\mathbf{k}) = -\frac{\chi}{\chi^2+(\mathbf{k}\cdot\mathbf{u})^2}\overline{\mathbf{u}'\mathbf{u}'^*}(\mathbf{k})\cdot\nabla_H\overline{\tau} \tag{A5}$$

In subsequent computations, we assume that the k-space density $\overline{\mathbf{u}'\mathbf{u}'^*}(\mathbf{k})$ is isotropic and thus:

$$\overline{\mathbf{u}'\mathbf{u}'^*}(\mathbf{k}) = \frac{1}{2}\overline{|\mathbf{u}'|^2}(k)\boldsymbol{\delta} \tag{A6}$$

Next, carrying out the integration described in (A2) with (A5), using (A6) with accuracy up to the main order in the ratio MKE/EKE, we obtain the following expression for the spectrum of the horizontal buoyancy flux:

$$\mathbf{F}_H(k) = -\chi^{-1}E(k)\nabla_H\overline{\tau} \tag{A7}$$

where $E(k) = \pi k\overline{|\mathbf{u}'|^2}(k)$ is the energy spectrum. Integrating (A7) over k and assuming that the shapes of the spectra are similar, we derive that:

$$\mathbf{F}_H = -\kappa_M\nabla_H\overline{\tau}, \qquad \kappa_M = r_d K^{1/2} \tag{A8}$$

## Appendix B. Vertical flux

We begin by deriving the expression for the z-derivative of the vertical tracer flux:

$$\partial_z F_v = \overline{w' \partial_z \tau'} + \overline{\tau' \partial_z w'} \tag{B1}$$

which enters the mean tracer equation (49a). Notice that (B1) is contributed only by the a-geostrophic component $\mathbf{u}_a$ of the eddy velocity. In particular, the second term can be rewritten as follows:

$$\overline{\tau' \partial_z w'} = -\overline{\tau' \nabla_H \cdot \mathbf{u}'} = -\overline{\tau' \nabla_H \cdot \mathbf{u}_a} \tag{B2}$$

In order to compute this correlation function, we follow the procedure described in Appendix A and consider the corresponding density in k space where $\mathbf{u}_a = (\mathbf{k}/k) u_a$. Thus, we have

$$\operatorname{Re} \overline{\tau'^* \partial_z w'}(\mathbf{k}) = \operatorname{Re}\left(-ik \overline{u_a \tau'^*}(\mathbf{k})\right) = k \operatorname{Im} \overline{u_a \tau'^*}(\mathbf{k}) \tag{B3}$$

The relation between eddy geostrophic and a-geostrophic components in the ML differs from that in the adiabatic ocean, Eq.(10a) of CD5, by the sign of the dynamical viscosity $\nu$ In fact, as we showed in CD5, in the adiabatic regime the enstrophy cascade cannot occur since the non-linear interactions do not conserve enstrophy. For this reason, there is only an inverse kinetic energy cascade corresponding to a negative dynamical viscosity. Near the surface, enstrophy is conserved by non-linear interactions and this allows an enstrophy cascade which results in a positive turbulent viscosity. As a result, in the ML, in Eq.(10a) of CD5 we must change the sign of $\nu$. Using relation (10a) of CD5 with the opposite sign of $\nu$ and the continuity equation, we obtain:

$$\partial_z w' = -ik u_a = k f^{-1} [\mathbf{k} \cdot \mathbf{u} - i\nu] u_g \tag{B4}$$

where $\mathbf{u}_g$ is the geostrophic component of the eddy velocity. Next, we substitute (B4) and Eq.(8) into (B3) and take into account that for mesoscales $u_a \ll u_g$ and therefore

$$\mathbf{u}' \approx \mathbf{u}_g = \mathbf{n} \times \mathbf{e}_z u_g , \quad \mathbf{n} = \mathbf{k}/|\mathbf{k}| \tag{B5}$$

In addition, since we adopt a turbulent Prandtl number $\sigma_t = 1$, we have $\nu = \chi$. The result is as follows:

$$\operatorname{Re} \overline{\tau'^* \partial_z w'}(\mathbf{k}) = -2(\chi f)^{-1} \mathbf{u} \cdot \mathbf{k} \mathbf{k} \times \mathbf{e}_z \cdot \nabla_H \bar{\tau} \overline{|u_g|^2}(\mathbf{k}) \tag{B6}$$

To derive the corresponding relation for the spectra, we substitute **(B6)** into (A2). Taking into account that the average of the tensor $k_i k_j$ yields $\delta_{ij}|\mathbf{k}|^2/2$, where $|\mathbf{k}| \approx k_0 = r_d^{-1}$, and that the spectrum of $\overline{|u_g|^2}$ equals 2E(k), we obtain:

$$\overline{\tau'\partial_z w'}(k) = -2(\chi fr_d^2)^{-1}\mathbf{u}\times\mathbf{e}_z \cdot \nabla_H \overline{\tau}\, E(k) \tag{B7}$$

Assuming that the shape of the spectra in the right and left hand sides are similar, we integrate over k which reduces to a substitution of the spectra with the corresponding variables. In addition, we use the second relations in Eq.(7) and (A8) to obtain:

$$\overline{\tau'\partial_z w'} = -2\varphi_M \mathbf{u}\times\mathbf{e}_z \cdot \nabla_H \overline{\tau} \quad , \quad \varphi_M \equiv \frac{\kappa_M}{fr_d^2} \tag{B8}$$

Next, we compute the first term of (B1). To do so, we need the expressions for $w'(t,\mathbf{k})$ and $\tau'_z(t,\mathbf{k})$. The former function can be derived by integrating (B4). With accuracy to the main order in z, we obtain:

$$w' = zkf^{-1}[\mathbf{k}\cdot\mathbf{u} - i\chi]u_g, \qquad z\mathbf{u}(z) = \int_0^z \mathbf{u}(z')dz' \tag{B9}$$

since $\chi = r_d^{-1} K^{1/2}$ and $u_g \approx K^{1/2}$ are almost constant within the ML. Differentiating Eq.(10) under the same condition, we get:

$$\partial_z \tau' \approx i\chi^{-2}\mathbf{k}\cdot\overline{\mathbf{u}}_z(\mathbf{u}'\cdot\nabla_H \overline{\tau}) \tag{B10}$$

Using (B9,10) to compute $\mathrm{Re}\,\overline{w'\partial_z\tau'^*}(\mathbf{k})$ and using a procedure analogous to (B6-8), we derive:

$$\overline{w'\partial_z \tau'} = z\varphi_M (\mathbf{e}_z \times \overline{\mathbf{u}}_z)\cdot \nabla_H \overline{\tau} \tag{B11}$$

Summing (B11) and (B7) and substituting into (B1) and using Eq.(7), we get:

$$\frac{\partial F_V(\tau)}{\partial z} = \mathbf{u}_* \cdot \nabla_H \overline{\tau} \tag{B12}$$

$$\mathbf{u}_* = \kappa_M \mathbf{F}(z)$$

$$\frac{1}{2}\mathrm{fr}_d^2 \mathbf{F}(z) = \mathbf{e}_z \times (\mathbf{u} + \frac{1}{2}z\partial_z \mathbf{u}), \qquad \mathbf{u} \equiv \overline{\mathbf{u}} - \mathbf{u}_d \qquad (B13)$$

## Appendix C. Derivation of relation (29e)

In the spirit of the procedure presented in Appendix A, we begin with expressing the Fourier component $p'(\mathbf{k})$ in terms of $\mathbf{u}'(\mathbf{k}) \approx \mathbf{u}_g(\mathbf{k})$. Using the geostrophic relation in $\mathbf{k}$-space $\mathbf{u}_g = \mathrm{i} f^{-1} \mathbf{k} p'$ and (B5), we obtain:

$$p'(\mathbf{k}) = \mathrm{i} f k^{-1} u_g(\mathbf{k}) \qquad (C1)$$

Using this relation together with (B4), we obtain the relation:

$$\mathrm{Re}\,\overline{p'^* \partial_z w'}(\mathbf{k}) = -\nu \overline{|u_g|^2}(\mathbf{k}) \qquad (C2)$$

which yields the spectral relation:

$$\overline{p' \partial_z w'}(k) = -2\nu E(k) \qquad (C3)$$

We may use this result in both ML and ocean interior. In the first case $\nu$ is positive while in the second case it is negative. In addition, since we adopt the turbulent Prandtl number $\sigma_t = 1$, we have $\nu = \chi$. Integrating (C3) over k and using the second of (7) of the main text, we arrive at (29e).

## Appendix D. Eddy surface kinetic energy: analytic study

To uncover the largest contribution to the eddy surface kinetic energy, we present an analytic treatment of (31). To do so, we decompose the mean velocity into Ekman and geostrophic parts:

$$\overline{\mathbf{u}}(z) = \mathbf{u}_E(z) + \mathbf{u}_g(z), \quad f\mathbf{u}_g(z) = \rho_0^{-1} \mathbf{e}_z \times \nabla_H \overline{p}$$
$$f\mathbf{u}_E(z) = -\rho_0^{-1} \mathbf{e}_z \times \partial_z \boldsymbol{\tau}, \quad \boldsymbol{\tau} = \rho_0 \nu_{\mathrm{turb}} \partial_z \overline{\mathbf{u}} \qquad (D1)$$

The geostrophic velocity is then decomposed into surface and thermal wind parts:

$$\mathbf{u}_g(z) = \mathbf{u}_S + \mathbf{u}_{tw}(z), \quad f\mathbf{u}_S = \rho_0^{-1} \mathbf{e}_z \times \nabla_H \overline{p}_S, \quad \mathbf{u}_{tw}(z) = z\partial_z \mathbf{u}_g, \quad f\partial_z \mathbf{u}_g(z) = \mathbf{e}_z \times \nabla_H \overline{b} \qquad (D2)$$

where $\bar{p}_S$ is the mean surface pressure contributed by the atmospheric pressure and by the one due to the sea surface height (Griffies, 2004, sec.3.8.1). Substituting relations (D1-2) into (25) and then into (30), we obtain:

$$K = K_S + K_{tw} + K_\tau \tag{D3}$$

where:

$$K_S = CD\mathbf{u}_s \cdot \partial_z \mathbf{u}_g, \qquad \mathbf{u}_s = \mathbf{u}_S - \mathbf{u}_d \tag{D4}$$

$$K_{tw} = -\frac{1}{2}CD^2|\partial_z \mathbf{u}_g|^2, \qquad K_\tau = Cf^{-2}\rho_0^{-1}\boldsymbol{\tau} \cdot \nabla_H \bar{b}[1+O(\delta_E/D)] \tag{D5}$$

where $\delta_E$ is Ekman layer's thickness. Next, we estimate the terms in (D3). In general, the surface velocity exceeds its interior counterpart and $\mathbf{u}_d$ and thus we take $\mathbf{u}_s \approx \mathbf{u}_S$. Furthermore, near the surface, the direction of $\partial_z \mathbf{u}_g$ is close to that of $\mathbf{u}_S$ and we assume that within the depth D, the variation of $\bar{\mathbf{u}}$ is a fraction $\alpha$ of $\mathbf{u}_S$:

$$D\partial_z \mathbf{u}_g \approx \alpha \mathbf{u}_S \tag{D6}$$

where $\alpha < 1$ but not very small. Thus, from (D4) we derive that:

$$K_S \approx 2\alpha C K_{MKE}, \qquad K_{MKE} = \frac{1}{2}|\mathbf{u}_S|^2 \tag{D7}$$

where $K_{MKE}$ is the geostrophic component of the surface mean kinetic energy due to the surface pressure discussed earlier. Next, from (D5) and (D4), we obtain:

$$K_{tw} \approx -\alpha^2 C K_{MKE} \tag{D8}$$

which shows that $K_{tw}$ is negative and smaller than $K_S$ by the factor $\sim \alpha/2$. To estimate the last term in (D3), we use the previous relations, together with $|\boldsymbol{\tau}(0)| = \rho_0 u_*^2$ and obtain that:

$$|K_\tau|/K_S \sim u_*^2/fD|\mathbf{u}_S| \tag{D9}$$

With $u_* \sim 10^{-2} ms^{-1}$ and D~100m, we obtain that $|K_\tau|/K_S \sim 0.1$.

In conclusion, the dominating term in (D3) is $K_S$ given by (D5). Since C>10, while $\alpha$ is not very small, *the surface eddy kinetic energy exceeds the (geostrophic component of the) mean kinetic energy by almost an order of magnitude,* a result in agreement with the conclusions of Scharffenberg and Stammer (2010).

A further assessment of relation (D5) comes from the numerical simulations of Capet et al. (2008). From their Fig.10, one obtains that $|\mathbf{u}_S| \approx 0.1 \text{ms}^{-1}$ and $|\partial_z \mathbf{u}_g| \approx 6 \cdot 10^{-4} \text{s}^{-1}$, D=40m. Thus, from (D5) we obtain $\alpha \approx 0.24$ and with C=16, we obtain $K_S = 3.6 \cdot 10^{-2} \text{m}^2 \text{s}^{-2}$ which compares well with Capet et al. (2008) value of $2.53 \cdot 10^{-2} \text{m}^2 \text{s}^{-2}$.

# Figure Caption

**Fig.1a.** Map of the length scale $\ell = \min(r_d, L_R)$. The Rossby deformation radius $r_d$ is obtained by solving Eq.(35) while the Rhines scale is defined as $L_R = (\overline{U}/\beta)^{1/2}$ where $\beta = 2 \times 10^{-11} \text{m}^{-1}\text{s}^{1}$.

**Fig.1b** Vertical buoyancy flux Eqs.(21-23) at four different locations, California Current, Gulf Stream, ACC and Labrador Sea (monthly and annual averages; h is the ML depth).

**Figs.2a** Surface eddy kinetic energy from T/P data (Scharffenberg and Stammer, 2010)

**Fig.2b** Model surface eddy kinetic energy, Eqs.(31).

**Fig.2c** Same as Fig.2b without the deep ocean contribution in Eq.(31)

**Fig.2d** Same as Fig.2b with the deep ocean contribution only in Eq.(31)

**Fig.2e** Zonal average of the surface kinetic energy from the altimetry data of Fig.2a and from the present model

**Fig.3a**. z-profiles of the eddy kinetic energy K(z) (in units of its surface value, see Eqs.34-35) vs. WOCE data for several locations

**Fig.3b** Same as Fig.3a for different locations

**Fig.4a** Surface mesoscale diffusivity computed from Eq.(15b) with the model for $K(0)=K_s$ shown in Fig.2b; the length scale $\ell$ is given in Fig.1a and the function $f(\overline{u}, K) = 1$

**Fig.4b** The z-profile of the mesoscale diffusivity computed from Eq.(15b) with the model for K(z) from relations (34-35) and plotted in Figs.3. The values are in units of the surface values.

**Fig.4c** Maps of the mesoscale diffusivity at different depths.

**Figs.5** Vertical profiles of the dynamical tapering function T(z) defined in in Eq.(39) for the ACC and Labrador Sea during the maximum ML extents.

**Figs.6a** ACC. The x-components of the old and new residual velocities defined in Eq.(53).

**Fig.6b.** Same as in Fig.6a for the y- components.

**Fig.7a** ACC. Polar map of the surface eddy kinetic energies from the T/P data (top panel) and model. The intensities compare quite reasonably.

**Fig.7b** ACC. Polar maps of the mesoscale diffusivities: values at the surface and at three different depths.

**Fig.7c** ACC. Polar maps of the tapering function T(z) for January, July and annual average. The values are computed at the bottom of the ML.

**Fig.7d** ACC. The x-y components of the stream wise integrated stream functions Eq.(56a).

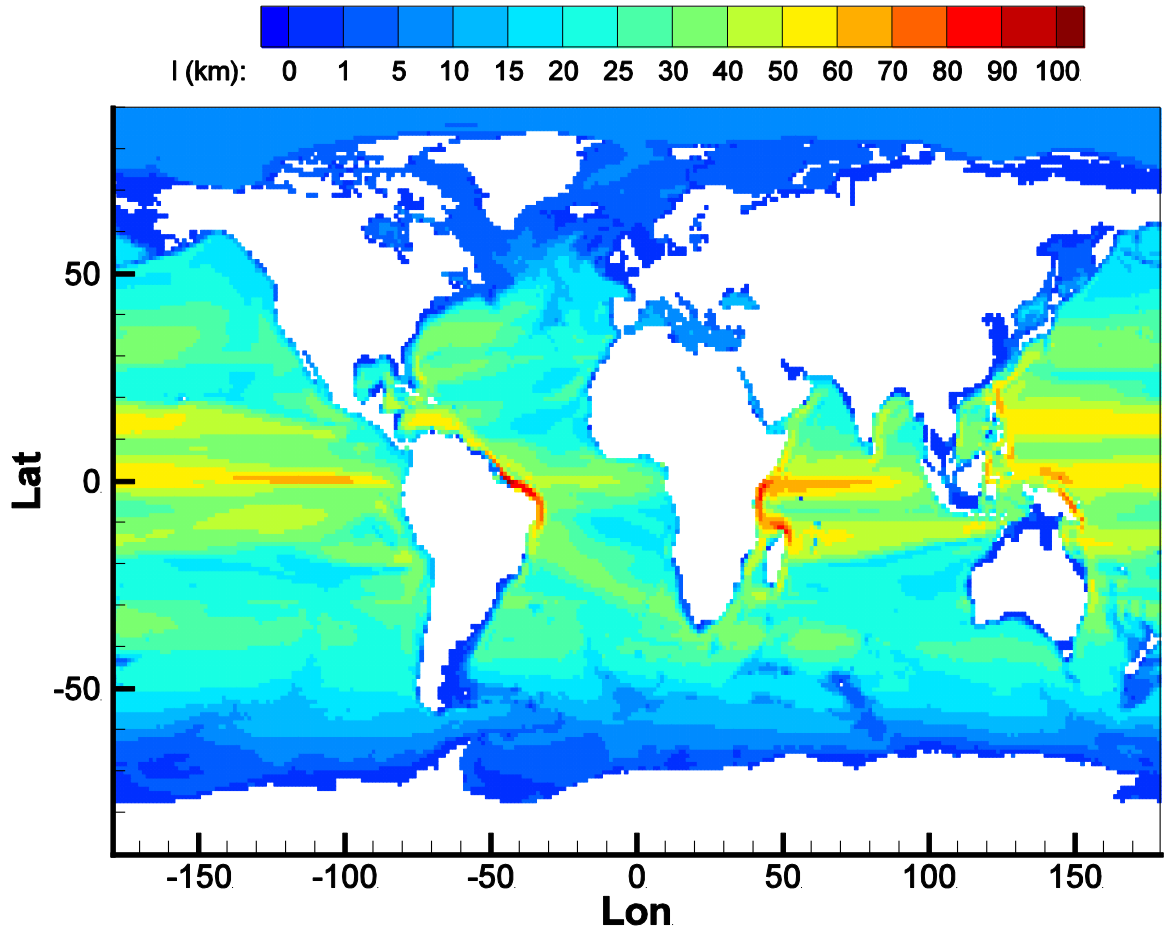

Fig. 1a

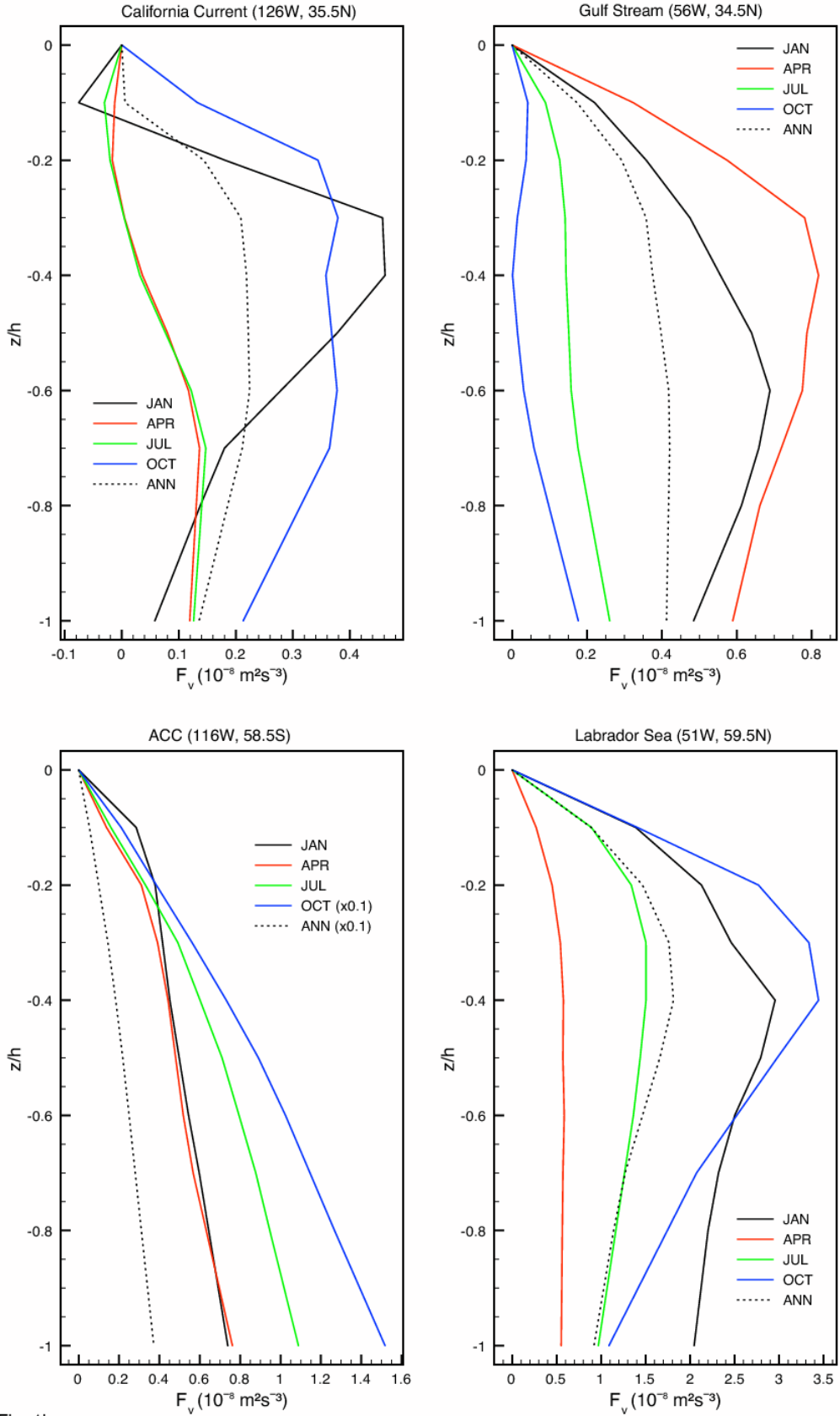

Fig. 1b

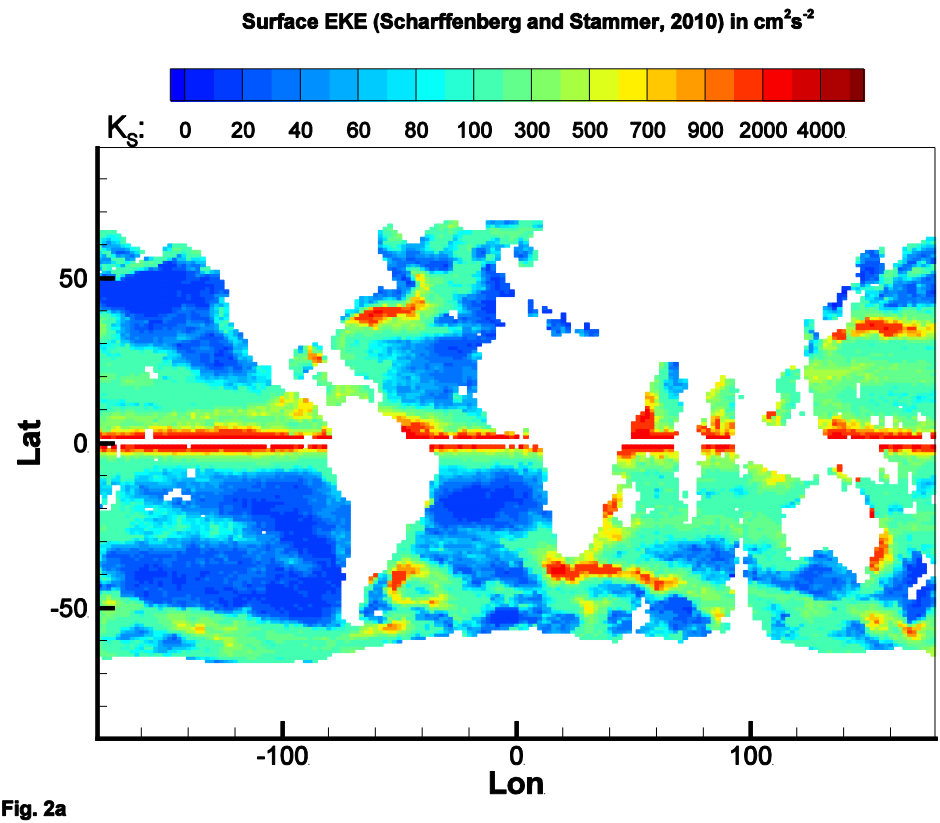

**Fig. 2a**

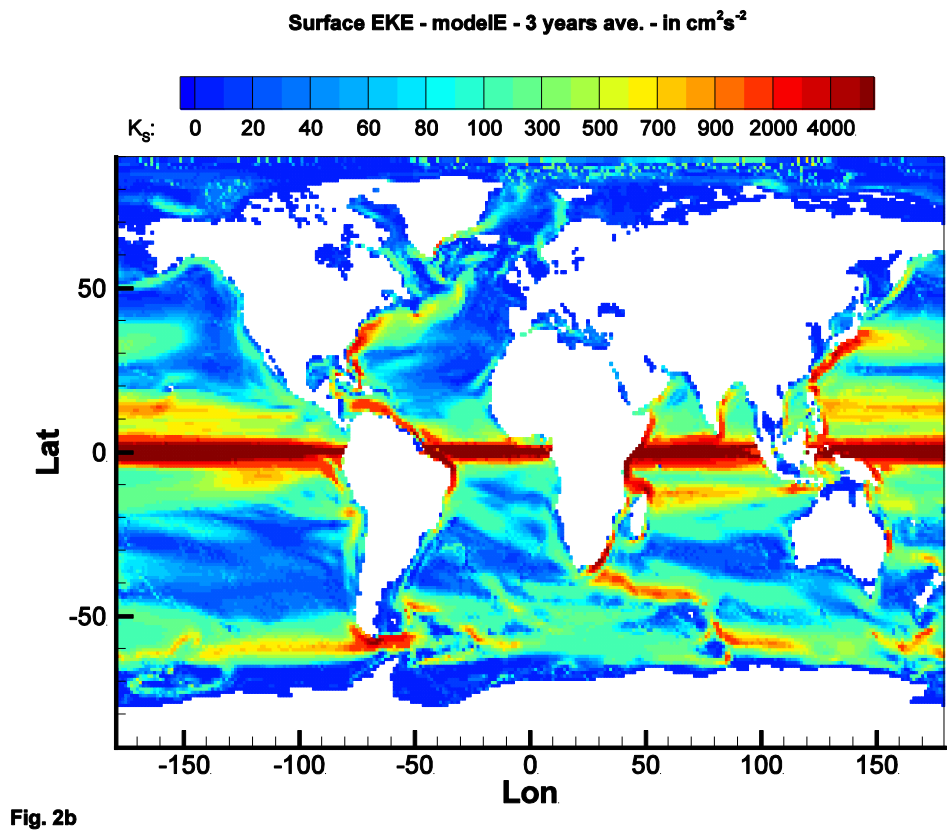

**Fig. 2b**

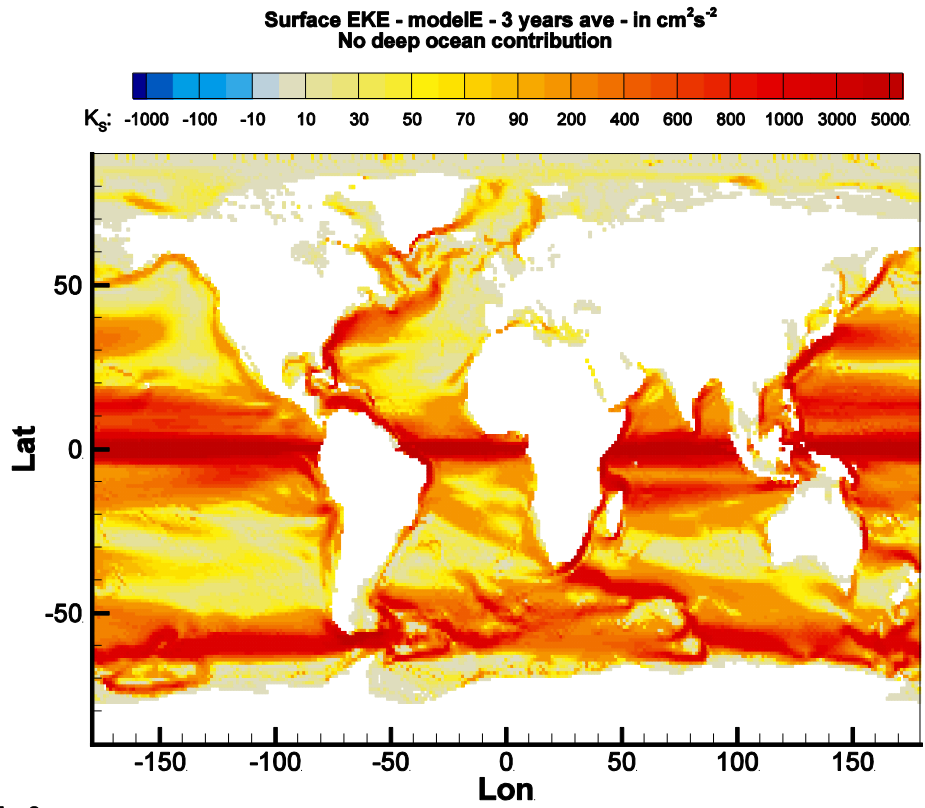

Fig. 2c

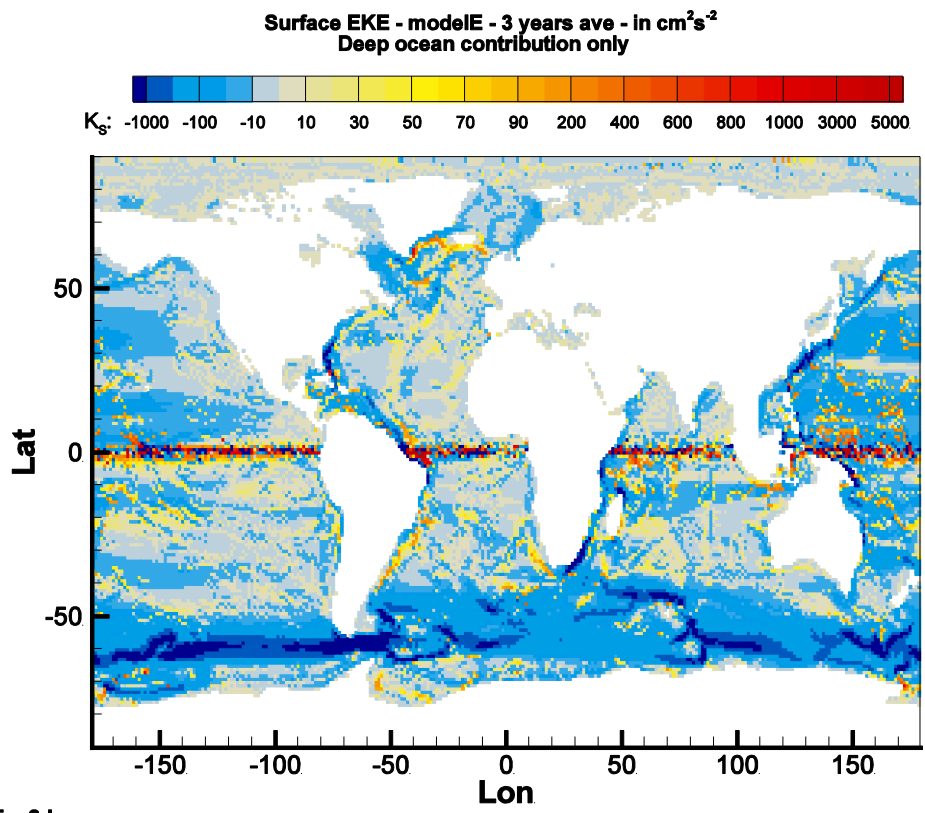

Fig. 2d

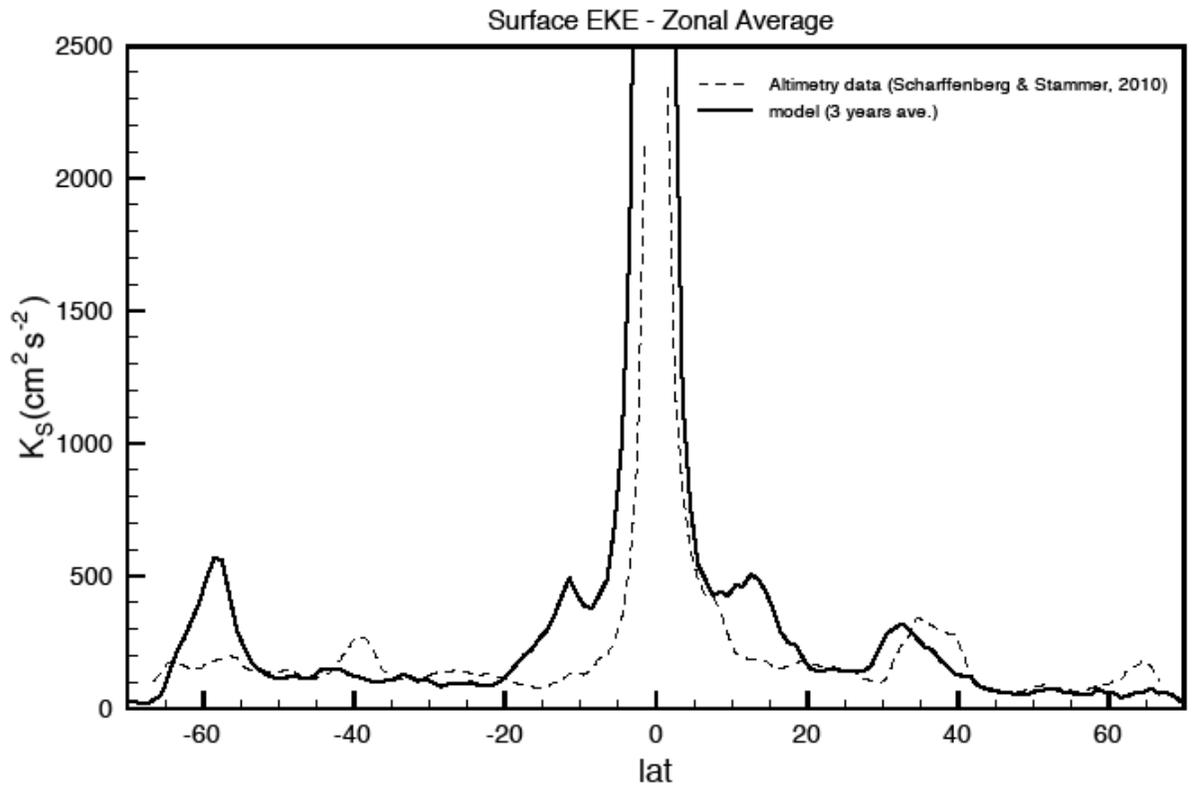

Fig. 2e

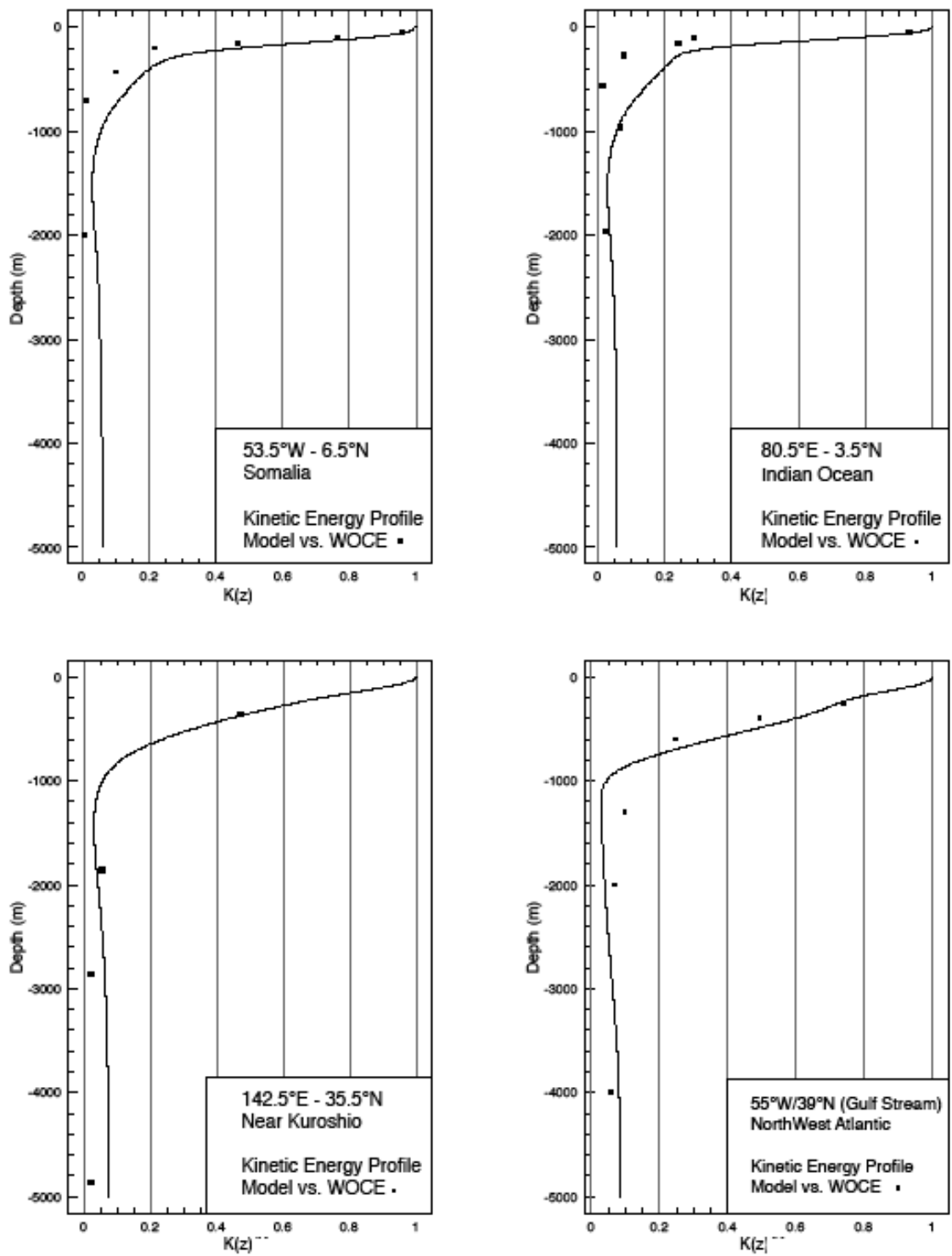

Fig. 3a

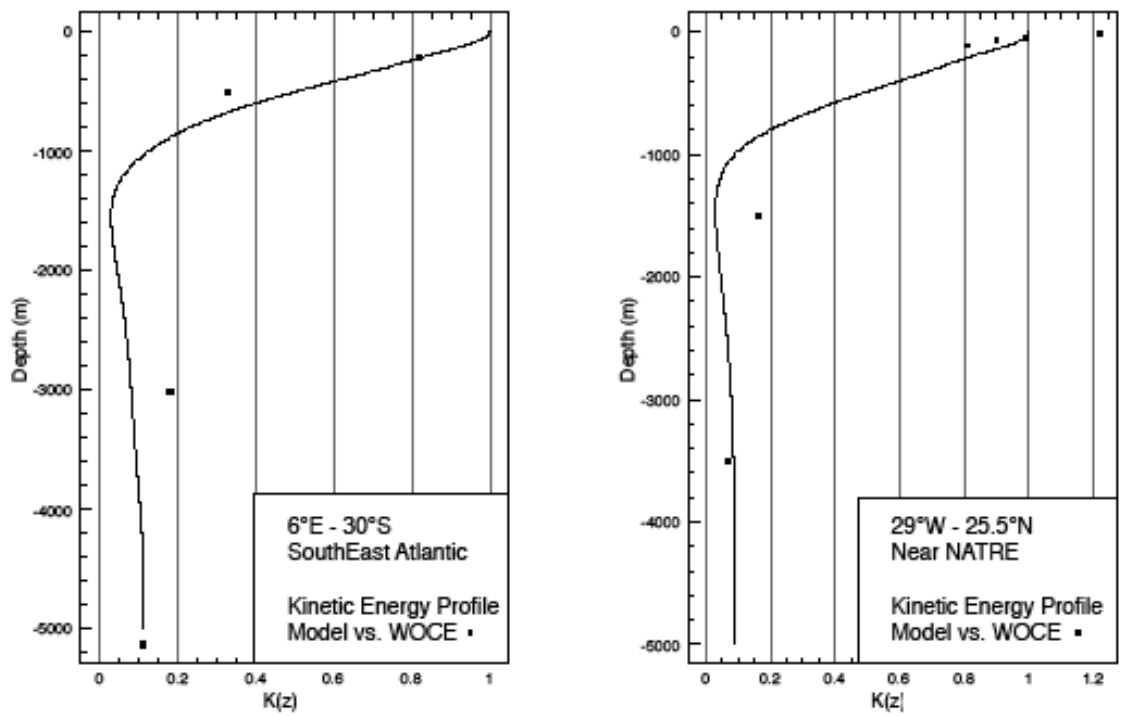

Fig. 3b

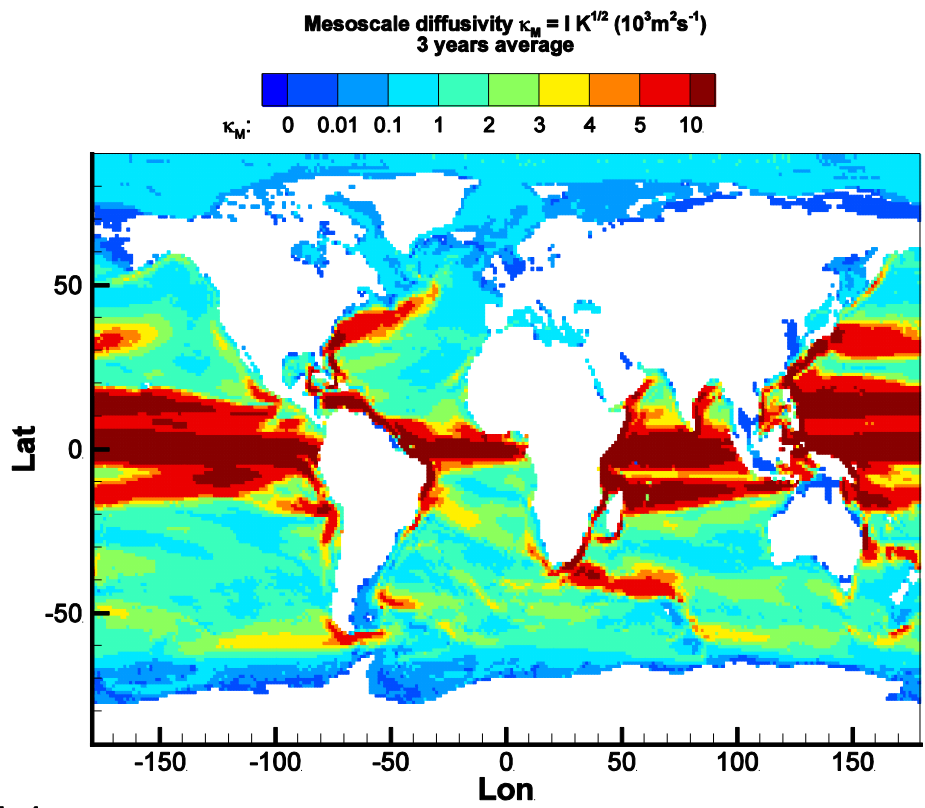

**Fig. 4a**

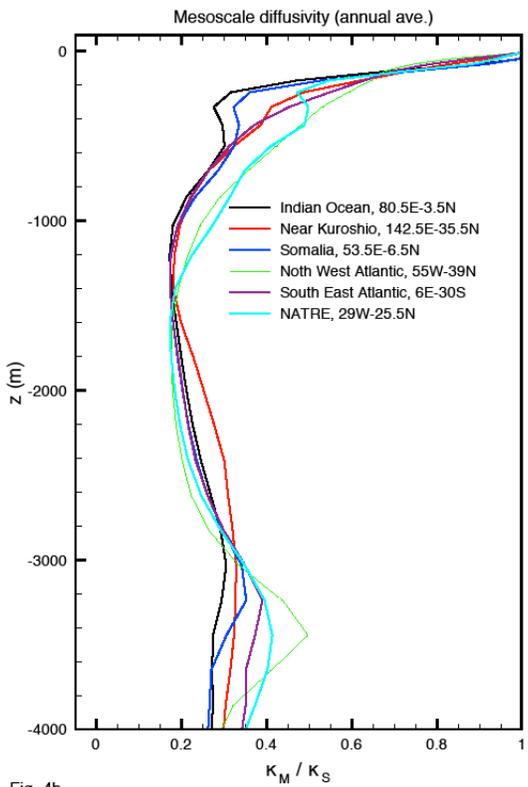

Fig. 4b

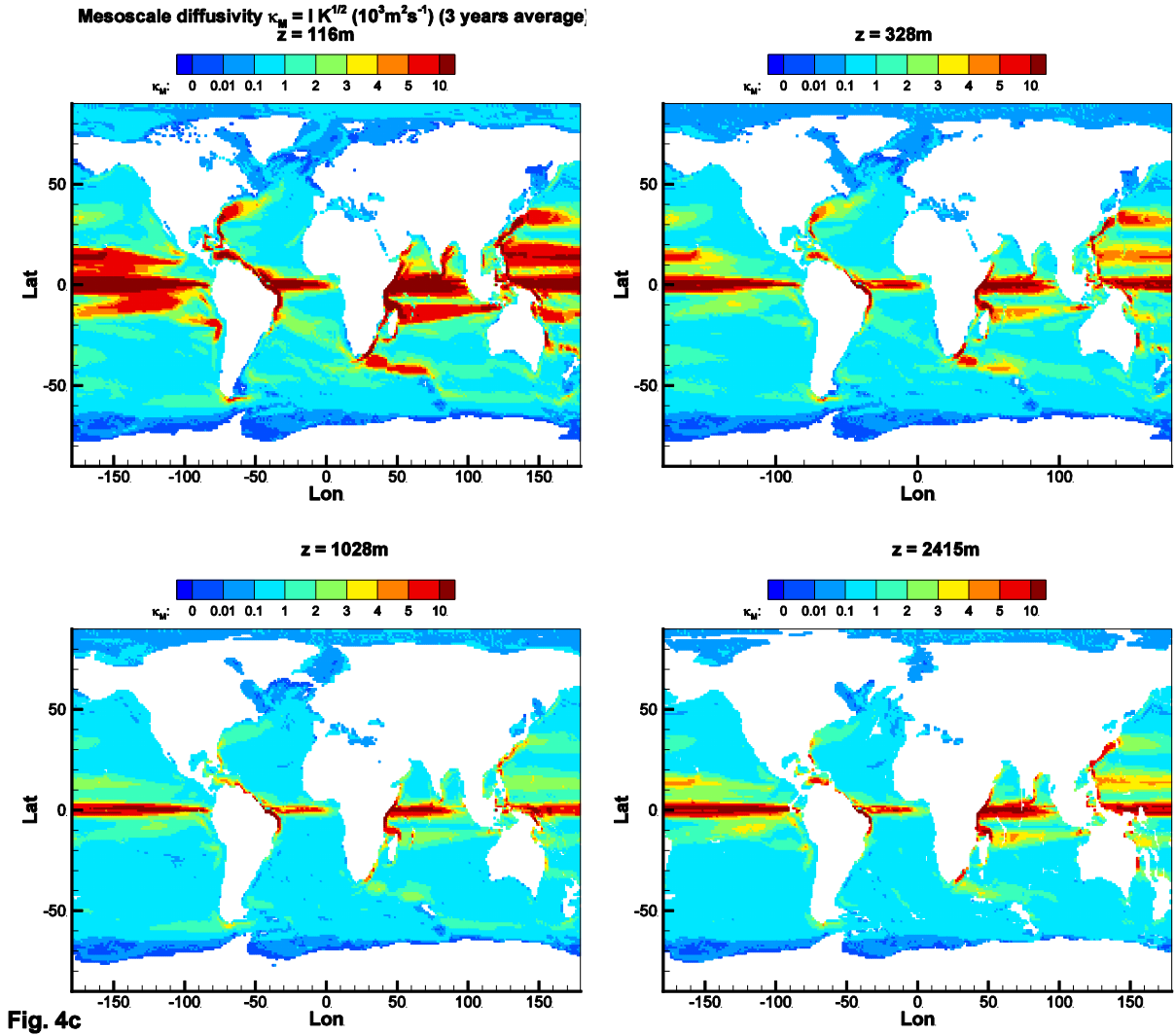

Fig. 4c

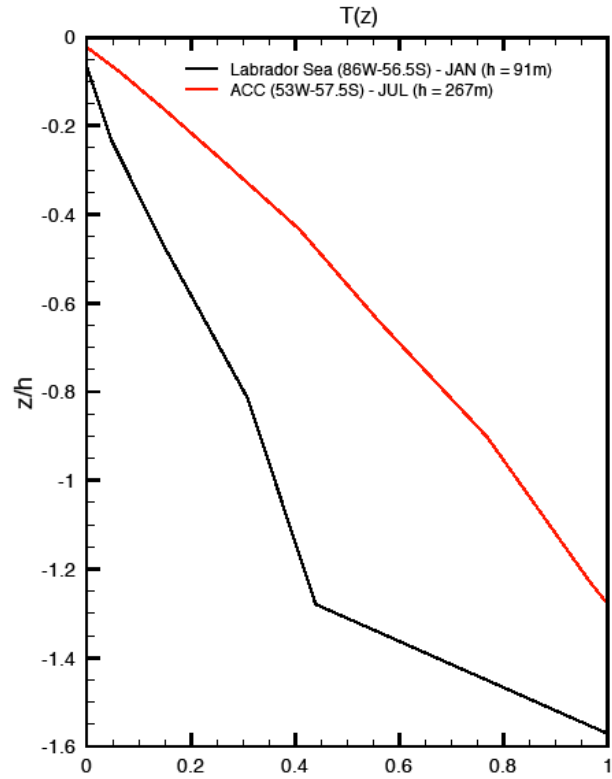

Fig. 5

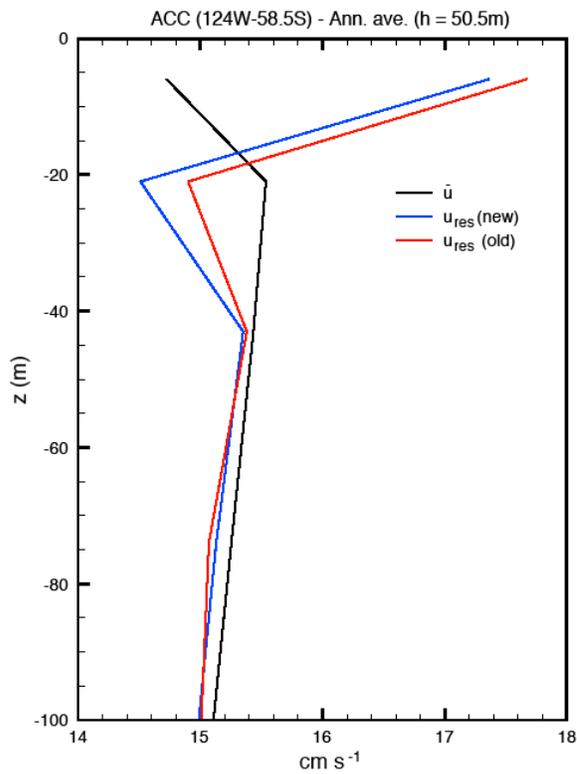

Fig. 6a

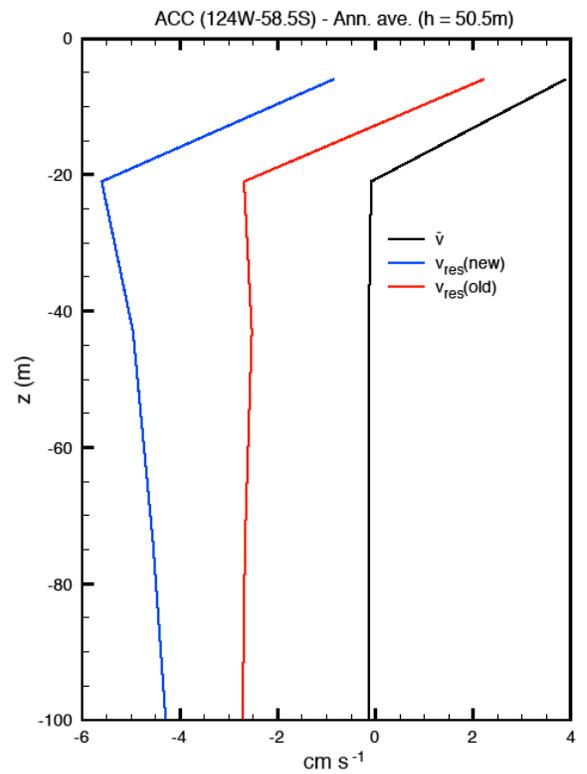

Fig. 6b

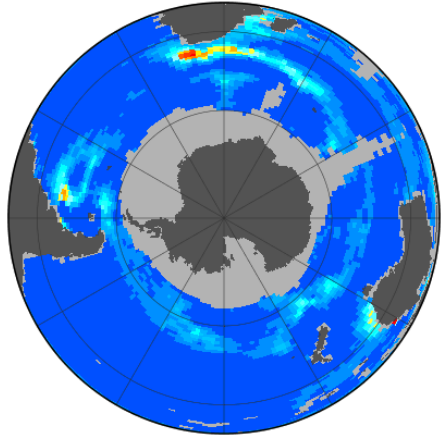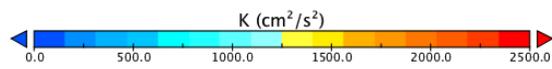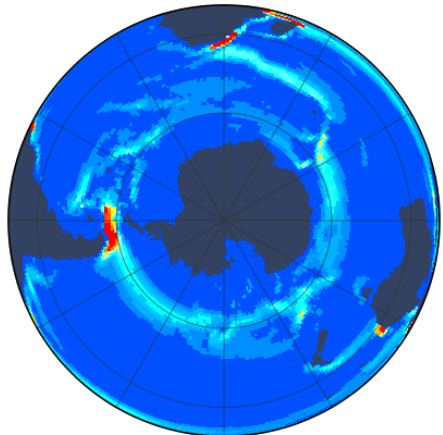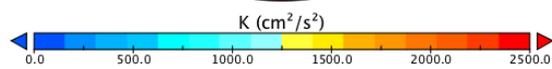

Fig. 7a

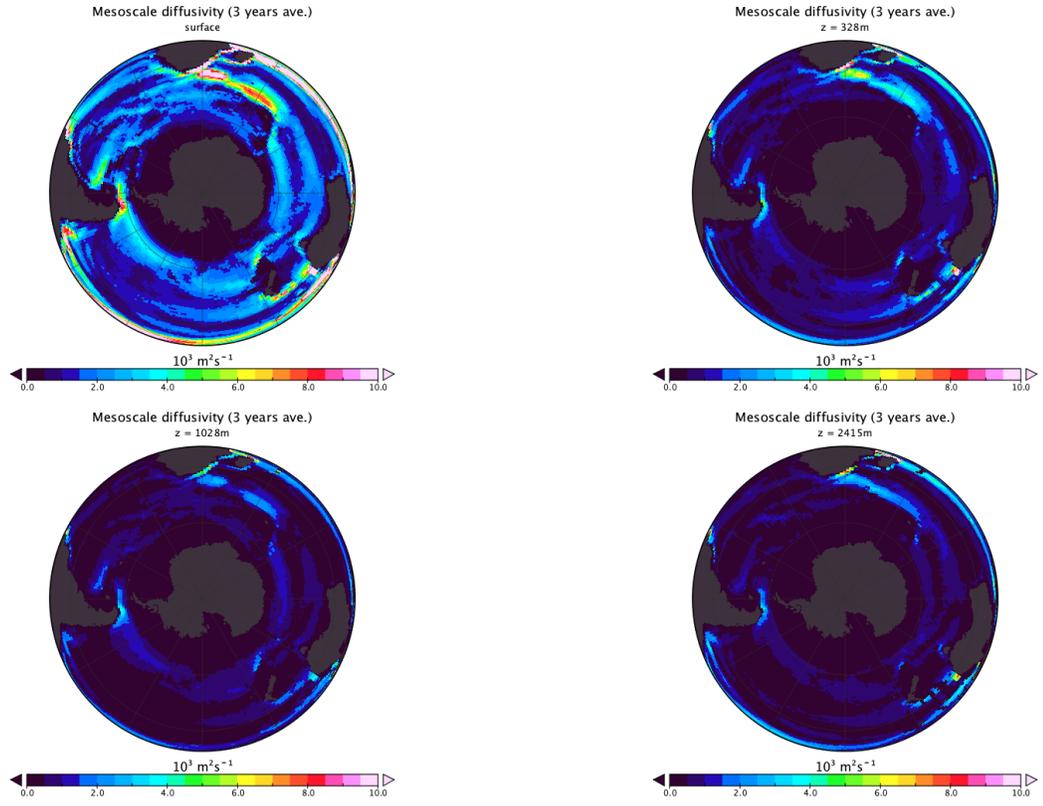

Fig. 7b

T(h) – January

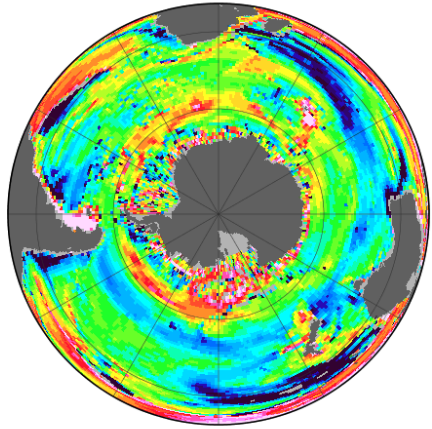

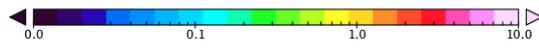

T(h) – July

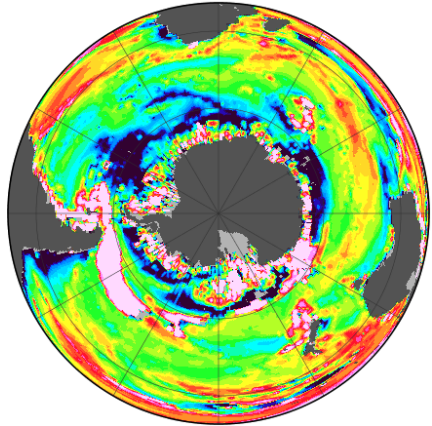

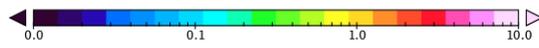

T(h) – Annual ave.

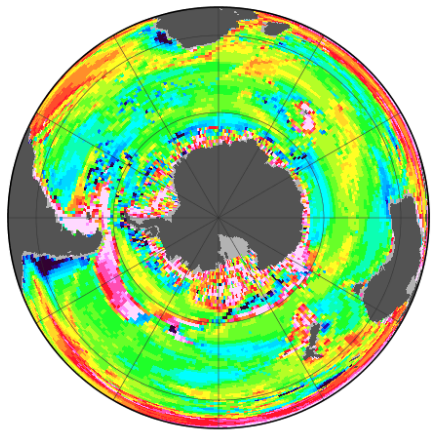

Fig. 7c

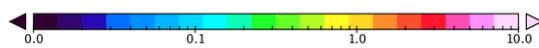

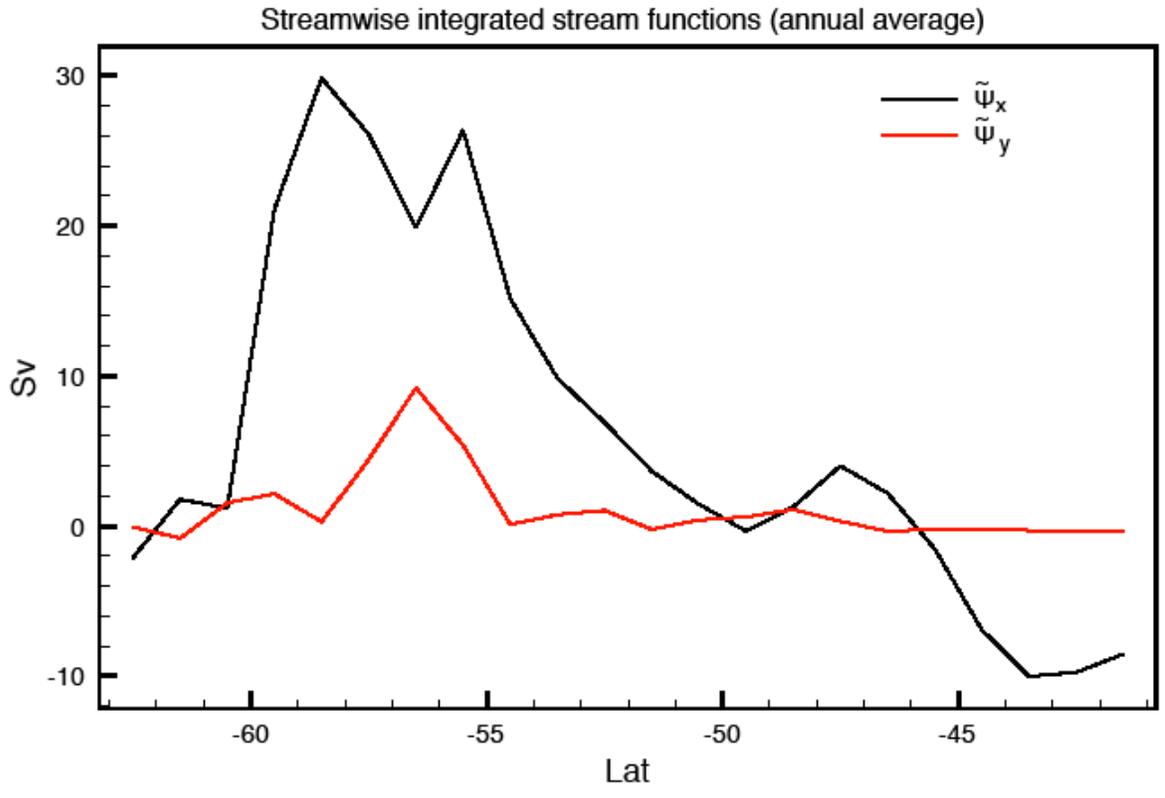

Fig. 7d

# Model Summary

**Dynamic equation for an arbitrary tracer** $\tau = T, S, C,$ etc

$$\partial_t \bar{\tau} + \underbrace{(\bar{\mathbf{U}} + \mathbf{U}^+ + \mathbf{U}_{**}) \cdot \nabla \bar{\tau} + \nabla \cdot \mathbf{F}_{diff}(\tau, b)}_{Mesoscales} + \underbrace{\nabla \cdot \mathbf{F}_{SM}}_{Sub-mesoscales} = -\partial_z F_{ss} + Q \tag{1}$$

Small scale turbulent mixing is represented by $F_{ss}$, sub-mesoscales are represented by $\mathbf{F}_{SM}$ and sources and sinks are represented by Q.

**Bolus Velocity derived from the stream function $\mathbf{U}^{++}= (\mathbf{u}^+, w^+)$**

$$\mathbf{u}^+ = \frac{\partial}{\partial z}[T(z)\,\mathbf{Y}(z)], \quad w^+ = -\nabla_H \cdot [T(z)\,\mathbf{Y}(z)], \quad \mathbf{Y}(z) = -\kappa_M \frac{\nabla_H \bar{b}(z)}{N_*^2} \tag{2}$$

**Tapering function (dimensionless):**

$$T(z) = -z \frac{N_*^2}{|\nabla_H \bar{b}|^2} \mathbf{F} \cdot \nabla_H \bar{b}, \quad T(0) = 0, \quad T(h_*) = 1 \tag{3}$$

**Definition of the depth $z = -h_*$:**

$$T(h_*) = 1, \quad N_*^2 \equiv N^2(z=-h_*) \tag{4}$$

The depth $h_*$ represents the boundary between the ML and the deep ocean.

**Function $\mathbf{F}(z)$ (dimension of inverse length)**

$$fr_d^2 \mathbf{F}(z) = \mathbf{u}(z) \times \mathbf{e}_z \tag{5a}$$

$$\mathbf{u}(z) \equiv z^{-1} \int_0^z \bar{\mathbf{u}}(z) dz + \bar{\mathbf{u}}(z) - 2\mathbf{u}_d \tag{5b}$$

is $\bar{\mathbf{u}}(z)$ the 2D mean velocity provided by the OGCM, $\mathbf{e}_z$ is the unit vertical vector (0,0,1) and $\mathbf{u}_d$ is eddy drift velocity, a barotropic quantity, given by:

**Eddy Drift Velocity:**

$$\mathbf{u}_d = <\bar{\mathbf{u}}> - \frac{1}{2} fr_d^2 \mathbf{e}_z \times (<\partial_z \mathbf{s}> - f^{-1}\boldsymbol{\beta}) \tag{6a}$$

$$<\bullet> \equiv \int_{-H}^{-h} \bullet \Gamma^{1/2}(z)dz / \int_{-H}^{-h} \Gamma^{1/2}(z)dz \tag{6b}$$

**Slope of the isopycnals s:**

$$\mathbf{s} = -N^{-2}\nabla_H \bar{b}, \quad N^2 = \bar{b}_z \tag{7}$$

$b = -g\rho_0^{-1}\rho$ is the buoyancy and is $\nabla_H$ is the 2D horizontal gradient

**Additional Bolus Velocity U∗∗:**

$$\mathbf{u}_{**} = \partial_z \mathbf{Y}_{**}, \quad w_{**} = -\nabla_H \cdot \mathbf{Y}_{**} \tag{8a}$$

$$\mathbf{Y}_{**} = -[1 - p(z)T(z)]\boldsymbol{\kappa}^{tr} \tag{8b}$$

$$p(z) \equiv \frac{N^2(z)}{N_*^2} \tag{8c}$$

$$\boldsymbol{\kappa}^{tr} = z\kappa_M[\mathbf{F} - (\frac{\mathbf{F} \cdot \nabla_H \bar{b}}{|\nabla_H \bar{b}|^2})\nabla_H \bar{b}] \tag{8d}$$

**Diffusive flux:**

$$\mathbf{F}_{diff}(\tau,b) = -\kappa_M \nabla_H \bar{\tau} - \kappa_M \boldsymbol{\Omega}(\tau,b)\frac{\partial \bar{\tau}}{\partial z} \tag{9a}$$

where the dimensionless vector $\boldsymbol{\Omega}(\tau,b)$ is given by:

$$\boldsymbol{\Omega}(\tau,b) = z[1-pT(z)]\mathbf{F}(z) - \frac{p}{N_*^2}T(z)^2 \nabla_H \bar{b} \tag{9b}$$

**Horizontal Mesoscale diffusivity:**

$$\kappa_M(z) = \ell K^{1/2}(z) \tag{10a}$$

**Length scale:** 
$$\ell = \min(r_d, L_R) \tag{10b}$$

$$r_d = \text{Rossby deformation radius}, \quad L_R = \text{Rhines scale} \tag{10c}$$

**K(z): Eddy kinetic energy K:**

$$K(z) = K(0)\Gamma(z) \tag{10d}$$

$$\Gamma(z) = (a_0^2 + |B_1(z)|^2)(1+a_0^2)^{-1} \quad , \quad a_0^2 \approx \overline{K}_{ML}/K(0) \tag{10e}$$

Here, $\overline{K}_{ML}$ is the mean kinetic energy $\overline{K}$ averaged over the ML. To compute the first baroclinic mode $B_1(z)$, one must solve the eigenvalue equation:

$$\partial_z(N^{-2}\partial_z B_1) + (r_d f)^{-2} B_1(z) = 0 \tag{10f}$$

with the boundary conditions $B_1(-h)=1, \partial_z B_1 = 0$ at z=- h, -H  The solution of (10f) also provides the Rossby radius $r_d$ needed in (5a) and (10c).

**Surface eddy kinetic energy**:

$$K(0) = \frac{C}{2fD} \int_{-D}^{0} dz\, |z|\, \mathbf{u}(z) \times \mathbf{e}_z \cdot \nabla_H \overline{b} \tag{11}$$